%% file: manuscript.tex
\documentclass[10pt,aps,prd,onecolumn,showpacs,showkeys,superscriptaddress,longbibliography]{revtex4-2}

\usepackage{amsmath,amssymb,amsfonts,amsthm}
\usepackage{mathtools}
\usepackage{bm}
\usepackage{graphicx}
\usepackage[hidelinks]{hyperref}
\usepackage{booktabs}
\usepackage{array}[=2016-10-06]
\usepackage{dcolumn}
\usepackage{xcolor}

\theoremstyle{definition}

\theoremstyle{remark}

\begin{document}

\nocite{*}

\title{
A Covariant Curvature-History Field Formulation for
State-Dependent Infinite-Derivative Gravity
}

\author{G.~M.~Rhythm}
\email{rhythm.research.astro@gmail.com}
\affiliation{School of Physics, Damghan University,
P.O.\ Box~3671641167, Damghan, Iran}

\date{July 15, 2026}

\begin{abstract}
We construct a covariant auxiliary-field formulation for state-dependent
infinite-derivative gravity.  Null-congruence curvature-history integrals
provide a natural motivation for state dependence, but they suffer from
variational ambiguities and loss of smooth metric dependence in the presence
of caustics and branch changes.  We replace the curve-dependent history
variable by a local scalar field \(C(x)\) satisfying a covariant hyperbolic
equation sourced by the Kretschmann invariant.  The history equation is
enforced at the level of the action by a conjugate auxiliary field
\(\chi(x)\).  Since the dynamical nonlocality scale depends on \(C(x)\), the
variation of the nonlocal form factor involves noncommuting operators; this
variation is evaluated using Duhamel's formula.  Treating \(C\) and \(\chi\)
as independent local fields, we derive the coupled field equations and show
that the stress-energy exchange between the nonlocal memory sector and the
auxiliary history sector cancels on shell.  The resulting local Noether
identity ensures covariant conservation of the combined memory-history
stress-energy tensor and compatibility with the contracted Bianchi identity.
\end{abstract}


\maketitle

\section{Introduction}
\label{sec:introduction}

Infinite-derivative modifications of gravitational dynamics provide a useful
effective framework for studying ultraviolet-softened gravitational interactions.
In these theories, local higher-derivative corrections are replaced by analytic
functions of the covariant d'Alembertian, typically chosen so that the
linearized propagator around simple backgrounds does not acquire additional
perturbative poles.  A representative form factor is governed by a fixed
nonlocality scale \(M_*\), which controls the onset of ultraviolet suppression
\cite{Biswas2012,Edholm2016,Modesto2012,Buoninfante2018}.  Such constructions
belong to the broader landscape of modified gravitational frameworks studied in
connection with high-curvature corrections, compact-object dynamics, collapse,
and stability analyses
\cite{Bhatti2020Stability,Asad2024Palatini,Yousaf2025AxialCollapse,Rehman2025Complexity},
although each of these applications requires a separate dynamical and spectral
analysis.

The present work addresses a narrower structural question.  In many situations
of physical interest, such as gravitational collapse, black-hole core dynamics,
or strongly curved cosmological evolution, it is natural to ask whether the
effective nonlocality scale should remain fixed or instead respond to the
evolving curvature state of the geometry.  This motivates replacing the
constant scale by a state-dependent one,
\begin{equation}
    M_*
    \longrightarrow
    M_{\rm eff}(C),
    \label{eq:scale_replacement_intro}
\end{equation}
where \(C(x)\) is a spacetime-dependent scalar field.  The central problem is
then not merely to choose a function \(M_{\rm eff}(C)\), but to define the
field \(C\) in a way that is covariant, smooth under metric variation, and
compatible with the contracted Bianchi identity.

In this formulation, \(C(x)\) is not introduced as an external regulator.  It
is a covariant curvature-history control field.  The field is taken to satisfy
a hyperbolic equation sourced by the Kretschmann invariant,
\begin{equation}
    \mathcal{K}
    =
    R_{\mu\nu\rho\sigma}R^{\mu\nu\rho\sigma}.
\end{equation}
With retarded boundary data, this equation allows \(C(x)\) to encode curvature
information from the causal past of \(x\).  The resulting scale
\(M_{\rm eff}(C)\) may therefore be interpreted as a state-dependent scale
controlling the strength of the nonlocal modification according to the
curvature history of the spacetime.

A direct way to motivate such a history variable is to integrate a curvature
scalar along past-directed null curves.  However, a single-ray null-congruence
definition is not well suited for use as a fundamental variable in a variational
action.  Null congruences can develop caustics and conjugate points, the
assignment of a unique ray to each spacetime point can become branch dependent,
and the metric variation of a line-integral functional generally produces
distributional support on the selected curves.  These difficulties do not make
all null-memory observables impossible, but they make a single-ray memory
functional unsuitable for the smooth local conservation mechanism developed in
this paper.

The main construction of this work replaces the curve-dependent history
variable by a local auxiliary-field system.  We introduce a scalar
curvature-history field \(C(x)\) and a conjugate auxiliary field \(\chi(x)\),
with \(\chi\) enforcing the curvature-history equation at the level of the
action.  This gives a covariant variational formulation in which the
state-dependent scale \(M_{\rm eff}(C)\) is not prescribed externally, but is
controlled by a dynamical field whose stress-energy contribution is included in
the metric variation.

The nonlocal memory sector contains a symmetric tensor field
\(\Delta_{\mu\nu}\) and an ordered state-dependent form factor.  Since
\(C(x)\) is spacetime dependent, multiplication by \(M_{\rm eff}^{-2}(C)\) does
not commute with the covariant d'Alembertian.  The variation of the form factor
therefore cannot be evaluated by an ordinary scalar chain rule.  We fix an
operator-ordering convention and use Duhamel's formula to evaluate the
noncommuting variation.  This identifies the source \(\Sigma_\Delta\) generated
by the \(C\)-dependence of the memory-sector action.

The central result is an on-shell conservation identity.  The memory sector and
the auxiliary curvature-history sector are not separately conserved when
\(M_{\rm eff}\) depends on \(C(x)\).  Instead, the memory-sector stress tensor
has a divergence proportional to \(\Sigma_\Delta\nabla_\nu C\), while the
auxiliary history sector produces the opposite term.  The exchange cancels on
shell:
\begin{equation}
    \nabla^\mu
    \left(
        T_{\mu\nu}^{(\Delta)}
        +
        T_{\mu\nu}^{(C)}
    \right)
    =
    0.
    \label{eq:intro_conservation_identity}
\end{equation}
Thus the state dependence of the nonlocal scale can be incorporated without
violating the local conservation requirement implied by the contracted Bianchi
identity.

The physical significance of the construction is that it provides a controlled
action-level mechanism for introducing curvature-dependent nonlocal scales in
modified gravity.  Such a mechanism is a prerequisite for studying whether
state-dependent nonlocality can affect high-curvature collapse, regular compact
objects, black-hole core dynamics, or effective dark-sector phenomenology.  In
response to the need for a more explicit physical interpretation, the present
paper also includes a representative application to black-hole core memory.  In
that application, the Kretschmann invariant sources the curvature-history field,
Ricci-flat black-hole regions remain active through the Weyl contribution to
\(\mathcal{K}\), and a scalar threshold can activate a nonzero memory phase
\(\Delta_{\mu\nu}\Delta^{\mu\nu}>0\) in the high-curvature core.  The activated
memory sector carries a variational stress tensor, remains compatible with the
Bianchi identity on shell, and provides a candidate stress-energy-carrying
encoding of collapse and curvature-history data inside the black-hole core.

This black-hole discussion is included to demonstrate the physical content of
the formalism, not to turn the present work into a complete black-hole
application paper.  A full treatment of core backreaction, stability, memory
capacity, singularity resolution, information release, or Page-curve physics
requires a separate analysis.  The present paper establishes the foundational
consistency step and gives a compact application-oriented demonstration of why
that step is physically useful.

The scope of the paper is therefore deliberately limited.  We prove neither
singularity resolution nor ghost freedom, and we do not derive a full
compact-object solution, a Page curve, a Hawking-radiation transfer mechanism,
or a dark-matter profile.  These questions require separate analyses of the
spectrum, background solutions, perturbative stability, causal boundary
conditions, and phenomenology.  The result established here is the foundational
consistency step: a state-dependent infinite-derivative scale can be coupled to
a covariant curvature-history field while preserving the on-shell stress-energy
conservation required by diffeomorphism invariance.

The paper is organized as follows.  Section~\ref{sec:geometric_setup} fixes the
geometric conventions, field content, dimensions, and boundary assumptions.
Section~\ref{sec:null_congruence_limitations} explains why a single-ray
null-congruence memory variable is not used as the fundamental action variable.
Section~\ref{sec:covariant_auxiliary_action} defines the covariant action and
the ordering convention for the state-dependent form factor.
Section~\ref{sec:auxiliary_history_equations} derives the auxiliary
curvature-history equations.  Section~\ref{sec:duhamel_variation} evaluates the
noncommuting variation of the form factor using Duhamel's formula.
Section~\ref{sec:memory_field_equation} derives the memory-field equation.
Section~\ref{sec:metric_variation} gives the metric variation and
stress-energy tensors.  Section~\ref{sec:conservation_identity} proves the
on-shell covariant conservation identity.  Section~\ref{sec:black_hole_core_application}
presents a representative black-hole core application, showing how curvature
history can activate a stress-energy-carrying memory phase and motivate a core
information-storage interpretation.  Section~\ref{sec:discussion} summarizes
the result and outlines future applications.


\section{Conventions, Field Content, and Boundary Data}
\label{sec:geometric_setup}

This section fixes the geometric conventions, field content, dimensional
assignments, and boundary assumptions used in the rest of the paper.  No
dynamical action is introduced here.  The action-level definition of the
memory sector and the auxiliary curvature-history sector is given in
Sec.~\ref{sec:covariant_auxiliary_action}.  The purpose of the present section
is only to specify the common mathematical setting in which the subsequent
variational construction is performed.

\subsection{Geometric conventions}
\label{subsec:spacetime_conventions}

Let \((\mathcal{M},g_{\mu\nu})\) be a four-dimensional time-oriented
Lorentzian manifold with metric signature
\begin{equation}
    (-,+,+,+).
\end{equation}
Greek indices \(\mu,\nu,\rho,\sigma,\ldots\) run over spacetime coordinates
\(0,1,2,3\).  The inverse metric is denoted by \(g^{\mu\nu}\), and indices are
raised and lowered with \(g_{\mu\nu}\) and \(g^{\mu\nu}\).

We use the Levi-Civita connection associated with \(g_{\mu\nu}\).  Thus
\begin{equation}
    \nabla_\rho g_{\mu\nu}=0,
    \qquad
    \Gamma^\rho_{\mu\nu}=\Gamma^\rho_{\nu\mu}.
\end{equation}
The Riemann tensor is defined by
\begin{equation}
    [\nabla_\mu,\nabla_\nu]V^\rho
    =
    R^\rho{}_{\sigma\mu\nu}V^\sigma ,
    \label{eq:riemann_convention}
\end{equation}
so that, in coordinates,
\begin{equation}
    R^\rho{}_{\sigma\mu\nu}
    =
    \partial_\mu\Gamma^\rho_{\nu\sigma}
    -
    \partial_\nu\Gamma^\rho_{\mu\sigma}
    +
    \Gamma^\rho_{\mu\lambda}\Gamma^\lambda_{\nu\sigma}
    -
    \Gamma^\rho_{\nu\lambda}\Gamma^\lambda_{\mu\sigma}.
\end{equation}
The Ricci tensor and Ricci scalar are
\begin{equation}
    R_{\mu\nu}=R^\rho{}_{\mu\rho\nu},
    \qquad
    R=g^{\mu\nu}R_{\mu\nu}.
\end{equation}
The Einstein tensor is
\begin{equation}
    G_{\mu\nu}
    =
    R_{\mu\nu}
    -
    \frac{1}{2}g_{\mu\nu}R,
\end{equation}
and satisfies the contracted Bianchi identity
\begin{equation}
    \nabla^\mu G_{\mu\nu}=0.
    \label{eq:bianchi_identity_section2}
\end{equation}

The covariant d'Alembertian acting on a scalar field is
\begin{equation}
    \Box
    \equiv
    g^{\mu\nu}\nabla_\mu\nabla_\nu .
    \label{eq:box_definition_section2}
\end{equation}
When \(\Box\) acts on tensor fields, the same notation denotes the covariant
rough Laplacian with the appropriate connection on the relevant tensor bundle.

\subsection{Curvature invariant used by the history field}
\label{subsec:curvature_history_invariant}

The curvature-history field introduced below is sourced by the Kretschmann
invariant
\begin{equation}
    \mathcal{K}
    \equiv
    R_{\mu\nu\rho\sigma}R^{\mu\nu\rho\sigma}.
    \label{eq:kretschmann_def_section2}
\end{equation}
This invariant is used because it detects tidal curvature even in Ricci-flat
regions.  For example, in the Schwarzschild exterior one has
\begin{equation}
    R_{\mu\nu}=0,
    \qquad
    R=0,
\end{equation}
whereas in general
\begin{equation}
    R_{\mu\nu\rho\sigma}R^{\mu\nu\rho\sigma}\neq 0.
\end{equation}
Thus a history variable sourced by \(\mathcal{K}\) responds to the full local
curvature strength, not only to Ricci curvature.  The corresponding
curvature-source scalar used later is
\begin{equation}
    \mathcal{I}[g]
    =
    \ell_*\mathcal{K},
    \label{eq:curvature_source_scalar_section2}
\end{equation}
where \(\ell_*\) is a fixed length scale.

\subsection{Field content}
\label{subsec:field_content_section2}

The fields used in the effective description are
\begin{equation}
    \left\{
        g_{\mu\nu},
        \Phi,
        \Delta_{\mu\nu},
        C,
        \chi
    \right\}.
    \label{eq:field_content_section2}
\end{equation}
Here \(g_{\mu\nu}\) is the spacetime metric and \(\Phi\) denotes an ordinary
matter field.  For definiteness, \(\Phi\) may be taken to be a minimally
coupled scalar, although the conservation argument below does not depend on
this particular choice.

The tensor field \(\Delta_{\mu\nu}\) is assumed symmetric,
\begin{equation}
    \Delta_{\mu\nu}=\Delta_{\nu\mu}.
    \label{eq:delta_symmetric_section2}
\end{equation}
It is treated as an effective geometric memory field.  Its dynamics are
specified by the memory-sector action in Sec.~\ref{sec:covariant_auxiliary_action}.

The scalar \(C\) is the curvature-history control field.  Its role is to
provide a covariant field-theoretic replacement for a curvature-history
variable that would otherwise be defined by an integral along selected null
curves.  In the action formulation, \(C\) is not imposed externally; it obeys a
hyperbolic equation sourced by \(\mathcal{K}\).  The conjugate scalar field
\(\chi\) enforces this curvature-history equation at the level of the action
and carries the corresponding variational response of the memory sector.

The state-dependent nonlocality scale used later is a positive scalar function
of \(C\),
\begin{equation}
    M_{\rm eff}^2=M_{\rm eff}^2(C),
    \qquad
    M_{\rm eff}^2(C)>0.
    \label{eq:Meff_positive_section2}
\end{equation}
The explicit operator ordering of the associated nonlocal form factor is not
defined in this section; it is fixed in Sec.~\ref{sec:covariant_auxiliary_action}
before the variation of the nonlocal operator is performed.

\subsection{Dimensional assignments}
\label{subsec:dimensional_assignments_section2}

We use natural units \(c=\hbar=1\).  Mass dimension is denoted by square
brackets.  Coordinates and derivatives have dimensions
\begin{equation}
    [x^\mu]=-1,
    \qquad
    [\partial_\mu]=1.
\end{equation}
The metric is taken to be dimensionless,
\begin{equation}
    [g_{\mu\nu}]=0.
\end{equation}
It follows that
\begin{equation}
    [\Gamma^\rho_{\mu\nu}]=1,
    \qquad
    [R^\rho{}_{\sigma\mu\nu}]=2,
    \qquad
    [R_{\mu\nu}]=2,
    \qquad
    [R]=2,
\end{equation}
and therefore
\begin{equation}
    [\mathcal{K}]=4.
    \label{eq:K_dimension_section2}
\end{equation}

The action is dimensionless,
\begin{equation}
    [S]=0,
\end{equation}
so the scalar Lagrangian density inside
\(\int d^4x\sqrt{-g}\,\mathcal{L}\) has dimension
\begin{equation}
    [\mathcal{L}]=4.
\end{equation}
The auxiliary fields are assigned
\begin{equation}
    [C]=1,
    \qquad
    [\chi]=1.
    \label{eq:C_chi_dimensions_section2}
\end{equation}
With this assignment,
\begin{equation}
    [\nabla_\mu\chi\nabla^\mu C]=4,
    \qquad
    [m_c^2\chi C]=4,
\end{equation}
provided
\begin{equation}
    [m_c]=1.
\end{equation}
Since \([\mathcal{K}]=4\), the coupling \(\ell_*\chi\mathcal{K}\) has
dimension four when
\begin{equation}
    [\ell_*]=-1.
\end{equation}
Thus \(\ell_*\) is a length scale.

For the memory tensor, the canonical two-derivative kinetic structure has the
schematic dimension
\begin{equation}
    [\Delta_{\mu\nu}\Box\Delta^{\mu\nu}]
    =
    2[\Delta]+2.
\end{equation}
Requiring this to be four gives
\begin{equation}
    [\Delta_{\mu\nu}]=1.
    \label{eq:Delta_dimension_section2}
\end{equation}
The mass scales appearing in the state-dependent nonlocality scale are assigned
\begin{equation}
    [M_*]=1,
    \qquad
    [C_*]=1,
    \qquad
    [M_{\rm eff}]=1.
\end{equation}

For reference, the main dimensional assignments are summarized in
Table~\ref{tab:dimension_assignments_section2}.

\begin{table}[t]
\centering
\begin{tabular}{c c}
\toprule
Quantity & Mass dimension \\
\midrule
\(x^\mu\) & \(-1\) \\
\(\partial_\mu,\nabla_\mu\) & \(1\) \\
\(g_{\mu\nu}\) & \(0\) \\
\(R^\rho{}_{\sigma\mu\nu},R_{\mu\nu},R\) & \(2\) \\
\(\mathcal{K}\) & \(4\) \\
\(\Phi\) & \(1\) \\
\(\Delta_{\mu\nu}\) & \(1\) \\
\(C,\chi\) & \(1\) \\
\(m_c,M_*,C_*,M_{\rm eff}\) & \(1\) \\
\(\ell_*\) & \(-1\) \\
\bottomrule
\end{tabular}
\caption{Mass-dimensional assignments used throughout the paper.}
\label{tab:dimension_assignments_section2}
\end{table}

\subsection{Regularity assumptions}
\label{subsec:regularity_assumptions_section2}

The variational derivations are carried out under smoothness assumptions strong
enough to justify integrations by parts and local tensor manipulations.  Unless
otherwise stated, we assume
\begin{equation}
    g_{\mu\nu}\in C^\infty(\mathcal{M}),
    \qquad
    \Phi\in C^\infty(\mathcal{M}),
    \qquad
    \Delta_{\mu\nu}\in C^\infty(\mathcal{M}),
    \qquad
    C,\chi\in C^\infty(\mathcal{M}).
    \label{eq:smoothness_assumptions_section2}
\end{equation}
Metric variations and field variations are taken either to have compact support
or to obey falloff conditions such that boundary terms vanish.  Equivalently,
one may supplement the action by the appropriate boundary terms and fix the
corresponding boundary data.  The present paper focuses on the bulk field
equations and the local Noether identities, so boundary contributions are not
displayed explicitly.

Under the assumptions in Eq.~\eqref{eq:smoothness_assumptions_section2},
\(\mathcal{K}\) is a smooth scalar wherever the metric is smooth.  Consequently,
the source \(\ell_*\mathcal{K}\) entering the curvature-history equation is also
smooth on such regions.  These assumptions are not intended to be optimal
functional-analytic hypotheses; they are adopted to isolate the covariance and
variational-consistency problem from the separate question of global existence
or maximal regularity of solutions.

\subsection{Boundary data and Green-function interpretation of the history field}
\label{subsec:boundary_data_section2}

The curvature-history equation derived in Sec.~\ref{sec:auxiliary_history_equations}
has the form
\begin{equation}
    P_C C
    =
    -\ell_*\mathcal{K},
    \qquad
    P_C\equiv \Box-m_c^2.
    \label{eq:history_operator_section2}
\end{equation}
As a local Euler--Lagrange equation, Eq.~\eqref{eq:history_operator_section2}
must be supplemented by boundary or initial data in order to select a particular
solution.  Since \(C\) is interpreted as a curvature-history field, the
physically relevant choice in a causal classical setting is the retarded
solution.

On a globally hyperbolic region where the retarded Green operator for \(P_C\)
exists, the retarded representative may be written formally as
\begin{equation}
    C(x)
    =
    C_{\rm hom}(x)
    -
    \ell_*
    \int_{\mathcal{M}} d^4y\,\sqrt{-g(y)}\,
    G_{\rm ret}(x,y)\mathcal{K}(y),
    \label{eq:C_retarded_solution_section2}
\end{equation}
where
\begin{equation}
    P_{C,x}G_{\rm ret}(x,y)
    =
    \frac{\delta^{(4)}(x-y)}{\sqrt{-g(y)}},
    \qquad
    {\rm supp}\,G_{\rm ret}(x,\cdot)\subseteq J^-(x),
    \label{eq:retarded_green_conditions_section2}
\end{equation}
and \(C_{\rm hom}\) satisfies
\begin{equation}
    P_C C_{\rm hom}=0.
\end{equation}
The homogeneous part \(C_{\rm hom}\) is fixed by the chosen initial data on a
Cauchy hypersurface.  Setting \(C_{\rm hom}=0\) corresponds to the case in which
the history field is generated entirely by the curvature source within the
chosen retarded domain.

The use of retarded boundary conditions is part of the physical interpretation
of \(C\) as a history variable.  The local variational identities derived below,
including the on-shell conservation law, do not require replacing \(C\) by the
explicit Green-function representation in Eq.~\eqref{eq:C_retarded_solution_section2}.
They follow from the local action and the equations of motion.  Thus the
retarded representation selects the causal solution branch, while the
covariant conservation mechanism is established at the level of the local
field formulation.

For the conjugate auxiliary field \(\chi\), the appropriate boundary data are
fixed by the variational problem and by the response equation derived from
variation with respect to \(C\).  In the conservation proof, \(\chi\) is treated
as an independent auxiliary field satisfying its own Euler--Lagrange equation.
No independent null-congruence prescription is assigned to \(\chi\).

\subsection{Role of this section}
\label{subsec:role_section2}

The assumptions above define the common geometric and functional setting for
the rest of the paper.  The curvature-history field \(C\), its conjugate
\(\chi\), and the memory tensor \(\Delta_{\mu\nu}\) have now been specified as
smooth covariant fields with consistent mass dimensions.  The dynamical action,
the operator-ordering convention for the state-dependent nonlocal form factor,
and the resulting Euler--Lagrange equations are introduced in the following
sections.

\section{Null-Congruence Memory and the Need for an Auxiliary Field}
\label{sec:null_congruence_limitations}

The curvature-history field used in this paper is motivated by the idea that a
state-dependent nonlocality scale should depend on accumulated curvature rather
than only on a local instantaneous invariant.  A geometrically natural first
attempt is to define such a history variable by integrating a curvature scalar
along past-directed null curves.  This section explains why this construction is
useful as motivation but is not adopted as the fundamental variable in the
action.

The issue is not that null-congruence observables are meaningless.  Rather, a
single-ray curvature-history functional is generally not a smooth, single-valued
functional of the metric on an arbitrary dynamical spacetime.  Caustics,
conjugate points, affine-normalization ambiguities, and distributional metric
variations make it unsuitable for the smooth variational mechanism developed in
later sections.  The auxiliary scalar \(C\) is introduced precisely to replace
this formal null-memory variable by a covariant field satisfying a local
hyperbolic equation.

\subsection{Formal null-history functional}
\label{subsec:null_history_candidate}

Let \(x\in\mathcal{M}\), and let \(\gamma_x(\lambda)\) be a past-directed null
geodesic ending at \(x\).  We take
\begin{equation}
    \gamma_x(0)=x,
    \qquad
    \lambda\leq 0,
    \label{eq:gamma_endpoint_section3}
\end{equation}
with tangent vector
\begin{equation}
    k^\mu
    =
    \frac{d\gamma_x^\mu}{d\lambda}.
    \label{eq:null_tangent_section3}
\end{equation}
The curve is affinely parametrized if
\begin{equation}
    k^\nu\nabla_\nu k^\mu=0,
    \label{eq:null_geodesic_equation_section3}
\end{equation}
and it is null if
\begin{equation}
    g_{\mu\nu}k^\mu k^\nu=0.
    \label{eq:null_condition_section3}
\end{equation}

A formal null-congruence curvature-history variable may then be written as
\begin{equation}
    C_{\rm null}(x)
    =
    \int_{\lambda_i(x)}^{0}
    \mathcal{I}[g]\bigl(\gamma_x(\lambda)\bigr)\,d\lambda ,
    \label{eq:C_null_precise_section3}
\end{equation}
where \(\lambda_i(x)<0\) denotes the lower endpoint of the chosen past segment.
The scalar \(\mathcal{I}[g]\) is a local curvature invariant.  For comparison
with the auxiliary-field construction used below, the natural choice is
\begin{equation}
    \mathcal{I}[g]
    =
    \ell_*\mathcal{K},
    \qquad
    \mathcal{K}
    =
    R_{\mu\nu\rho\sigma}R^{\mu\nu\rho\sigma}.
    \label{eq:null_memory_integrand_section3}
\end{equation}
With this choice, the null-history variable measures accumulated tidal
curvature along the selected past-directed null ray.

Equation~\eqref{eq:C_null_precise_section3} is not yet a complete definition.
An affine parameter is not unique: if \(\lambda\) is affine, then
\begin{equation}
    \lambda\mapsto a\lambda+b,
    \qquad
    a>0,
    \label{eq:affine_rescaling_section3}
\end{equation}
is also affine.  Since the integral in
Eq.~\eqref{eq:C_null_precise_section3} depends on the normalization of
\(d\lambda\), a physical prescription must fix this freedom.  For example, one
may introduce a future-directed unit timelike vector field \(u^\mu\) associated
with a family of observers and impose the endpoint normalization
\begin{equation}
    -u_\mu(x)k^\mu(x)=1.
    \label{eq:affine_normalization_section3}
\end{equation}
Alternatively, one may normalize \(k^\mu\) on an initial hypersurface or at a
specified boundary.  Each of these choices introduces additional structure
beyond the metric alone.  Thus a single-ray expression such as
Eq.~\eqref{eq:C_null_precise_section3} is not, by itself, a canonical scalar
functional of \(g_{\mu\nu}\).

The lower limit \(\lambda_i(x)\) also requires specification.  It may represent
the intersection of the null ray with an initial Cauchy hypersurface, a finite
cutoff surface, or a past boundary of the spacetime region under consideration.
Different choices correspond to different history prescriptions.  This
dependence is not necessarily problematic for an observable, but it is
undesirable for a fundamental field variable placed inside a local action unless
the corresponding boundary data are specified as part of the variational
problem.

For these reasons, Eq.~\eqref{eq:C_null_precise_section3} will be treated as a
motivating expression rather than as the definition of the field used in the
action.  The action-based formulation below replaces \(C_{\rm null}\) by a
smooth scalar field \(C\) satisfying a covariant hyperbolic equation with
retarded boundary data when a causal history interpretation is desired.

\subsection{Caustics and loss of smooth dependence on the endpoint}
\label{subsec:caustics_section3}

A second difficulty is that the assignment
\begin{equation}
    x\mapsto \gamma_x
    \label{eq:endpoint_to_curve_map_section3}
\end{equation}
is not generally smooth or unique on a curved spacetime.  Consider a null
geodesic congruence with tangent \(k^\mu\).  Let \(\xi^\mu\) be a deviation
vector connecting neighboring rays.  The relative acceleration of neighboring
geodesics is governed by the geodesic-deviation equation
\begin{equation}
    \frac{D^2\xi^\mu}{d\lambda^2}
    =
    R^\mu{}_{\nu\rho\sigma}
    k^\nu k^\rho \xi^\sigma ,
    \qquad
    \frac{D}{d\lambda}=k^\mu\nabla_\mu .
    \label{eq:geodesic_deviation_section3}
\end{equation}
Thus curvature controls the focusing and defocusing of nearby null rays.

To describe focusing, introduce an auxiliary null vector \(l^\mu\) satisfying
\begin{equation}
    k_\mu l^\mu=-1,
    \qquad
    l_\mu l^\mu=0.
\end{equation}
The screen-space metric transverse to the null directions is
\begin{equation}
    q_{\mu\nu}
    =
    g_{\mu\nu}
    +
    k_\mu l_\nu
    +
    l_\mu k_\nu .
    \label{eq:screen_metric_section3}
\end{equation}
For two independent transverse deviation vectors \(\xi^\mu_{(1)}\) and
\(\xi^\mu_{(2)}\), the infinitesimal cross-sectional area of the congruence is
schematically
\begin{equation}
    A(\lambda)
    =
    \sqrt{
    \det\!\left[
    q_{\mu\nu}\xi^\mu_{(a)}\xi^\nu_{(b)}
    \right]
    },
    \qquad
    a,b=1,2.
    \label{eq:cross_sectional_area_section3}
\end{equation}
The null expansion is
\begin{equation}
    \theta
    =
    \nabla_\mu k^\mu ,
    \label{eq:null_expansion_section3}
\end{equation}
after the usual projection onto the screen space.  The area evolves according
to
\begin{equation}
    \frac{dA}{d\lambda}
    =
    \theta A.
    \label{eq:area_evolution_section3}
\end{equation}
For a hypersurface-orthogonal null congruence, the Raychaudhuri equation takes
the form
\begin{equation}
    \frac{d\theta}{d\lambda}
    =
    -\frac12\theta^2
    -
    \sigma_{\mu\nu}\sigma^{\mu\nu}
    -
    R_{\mu\nu}k^\mu k^\nu ,
    \label{eq:null_raychaudhuri_section3}
\end{equation}
where \(\sigma_{\mu\nu}\) is the shear tensor.  More generally, a twist term is
present, but it vanishes for hypersurface-orthogonal light-cone congruences.

A caustic forms when neighboring rays intersect and the cross-sectional area
vanishes:
\begin{equation}
    A(\lambda_c)=0
    \label{eq:caustic_condition_section3}
\end{equation}
for some finite affine parameter \(\lambda_c\).  At such a point the
congruence no longer provides a smooth one-to-one parametrization of the null
surface.  Consequently, even if \(\mathcal{I}[g]\) is a smooth scalar, the
map
\begin{equation}
    x\mapsto C_{\rm null}(x)
\end{equation}
may fail to be smooth because the underlying assignment of a unique past null
ray fails to be smooth.

This is a structural obstruction to using \(C_{\rm null}\) as a fundamental
field variable in a local variational action.  A field appearing in the action
should have a controlled metric variation.  A single-null-ray history
functional can lose this property precisely in the regions where curvature
focusing is important.

\subsection{Conjugate points and branch dependence}
\label{subsec:nonuniqueness_section3}

The same obstruction can be expressed in terms of the exponential map.  Let
\(p\in\mathcal{M}\), and let \(\exp_p\) denote the exponential map.  A null
geodesic emitted from \(p\) can be written locally as
\begin{equation}
    \gamma(\lambda)
    =
    \exp_p(\lambda k),
    \label{eq:exponential_map_section3}
\end{equation}
where \(k^\mu\) is a null tangent vector at \(p\).  The null-geodesic
coordinates defined by this map are locally regular only when the differential
of the exponential map has full rank,
\begin{equation}
    \det(D\exp_p)\neq 0 .
    \label{eq:regular_exp_map_section3}
\end{equation}
At conjugate points,
\begin{equation}
    \det(D\exp_p)=0 .
    \label{eq:singular_exp_map_section3}
\end{equation}
The null-ray parametrization then becomes singular.

After conjugate points, several distinct null geodesics may connect the same
causal region to the endpoint \(x\).  A single-ray history variable would then
require a branch choice,
\begin{equation}
    \gamma_x
    \in
    \left\{
    \gamma_x^{(1)},\gamma_x^{(2)},\ldots
    \right\}.
    \label{eq:branch_choice_section3}
\end{equation}
Unless this branch is selected by an additional covariant prescription,
\(C_{\rm null}(x)\) is not uniquely defined.  If a noncovariant prescription is
used, the resulting quantity is not suitable as a fundamental scalar in a
diffeomorphism-invariant action.

One could instead average over all branches or integrate over an entire light
cone.  Such constructions may be meaningful in other contexts, but they define
a different nonlocal object and require their own measure, regularization, and
variational analysis.  The present work follows a different route: it replaces
the branch-dependent null integral by a local auxiliary scalar whose dynamics
are fixed by a covariant equation.

\subsection{Metric variation and distributional support}
\label{subsec:metric_variation_distributional_section3}

The most direct difficulty appears when a null-history variable is inserted
inside an action.  Suppose, for illustration, that the action contains a term
of the schematic form
\begin{equation}
    S_{\rm hist}
    =
    \int_{\mathcal{M}} d^4x\sqrt{-g(x)}\,
    \mathcal{L}\!\left(C_{\rm null}(x)\right).
    \label{eq:S_hist_null_section3}
\end{equation}
The metric variation contains
\begin{equation}
    \delta S_{\rm hist}
    \supset
    \int_{\mathcal{M}} d^4x\sqrt{-g(x)}\,
    \frac{\partial\mathcal{L}}{\partial C_{\rm null}}(x)
    \,
    \delta C_{\rm null}(x).
    \label{eq:delta_S_hist_chain_section3}
\end{equation}
From Eq.~\eqref{eq:C_null_precise_section3},
\begin{align}
    \delta C_{\rm null}(x)
    &=
    \delta
    \int_{\lambda_i(x)}^{0}
    \mathcal{I}[g]\bigl(\gamma_x(\lambda)\bigr)\,d\lambda
    \nonumber \\
    &=
    \int_{\lambda_i(x)}^{0}
    \delta\mathcal{I}[g]\bigl(\gamma_x(\lambda)\bigr)\,d\lambda
    +
    \int_{\lambda_i(x)}^{0}
    \nabla_\alpha\mathcal{I}[g]\bigl(\gamma_x(\lambda)\bigr)
    \,
    \delta\gamma_x^\alpha(\lambda)\,d\lambda
    +
    \delta_{\rm end}C_{\rm null}(x).
    \label{eq:delta_C_null_split_section3}
\end{align}
The first term varies the curvature scalar at fixed curve.  The second term
comes from the metric dependence of the geodesic itself.  The last term denotes
possible endpoint contributions if the lower endpoint \(\lambda_i(x)\), the
initial hypersurface, or the normalization prescription varies with the metric.

The first term already displays the distributional structure.  Since
\(\mathcal{I}[g]\) is local in the metric and its derivatives, its functional
variation can be written schematically as
\begin{equation}
    \delta\mathcal{I}[g]\bigl(\gamma_x(\lambda)\bigr)
    =
    \int_{\mathcal{M}} d^4z\,
    \frac{
    \delta\mathcal{I}\bigl(\gamma_x(\lambda)\bigr)
    }{
    \delta g^{\mu\nu}(z)
    }
    \delta g^{\mu\nu}(z).
    \label{eq:delta_I_functional_section3}
\end{equation}
For a local curvature scalar, the kernel has support only when
\begin{equation}
    z=\gamma_x(\lambda).
    \label{eq:support_condition_section3}
\end{equation}
Therefore it contains derivatives of covariant Dirac distributions of the
schematic form
\begin{equation}
    \frac{
    \delta\mathcal{I}\bigl(\gamma_x(\lambda)\bigr)
    }{
    \delta g^{\mu\nu}(z)
    }
    \sim
    \mathcal{D}_{\mu\nu}
    \left[
    \frac{
    \delta^{(4)}\!\left(z-\gamma_x(\lambda)\right)
    }{
    \sqrt{-g(z)}
    }
    \right],
    \label{eq:distributional_kernel_section3}
\end{equation}
where \(\mathcal{D}_{\mu\nu}\) is a differential operator determined by the
chosen invariant \(\mathcal{I}[g]\).  Substitution into
Eq.~\eqref{eq:delta_S_hist_chain_section3} gives line-supported terms in the
functional derivative of the action:
\begin{align}
    \delta S_{\rm hist}
    &\supset
    \int_{\mathcal{M}} d^4z\,
    \delta g^{\mu\nu}(z)
    \int_{\mathcal{M}} d^4x\sqrt{-g(x)}
    \frac{\partial\mathcal{L}}{\partial C_{\rm null}}(x)
    \nonumber \\
    &\hspace{2.0cm}\times
    \int_{\lambda_i(x)}^{0} d\lambda\,
    \mathcal{D}_{\mu\nu}
    \left[
    \frac{
    \delta^{(4)}\!\left(z-\gamma_x(\lambda)\right)
    }{
    \sqrt{-g(z)}
    }
    \right].
    \label{eq:delta_S_distributional_section3}
\end{align}
Thus the stress tensor derived from \(S_{\rm hist}\) is not generically a
smooth tensor field.  It contains contributions supported on the selected null
curves.

The second term in Eq.~\eqref{eq:delta_C_null_split_section3} introduces a
separate complication.  Varying the geodesic equation gives a forced Jacobi
equation for the curve variation:
\begin{equation}
    \frac{D^2}{d\lambda^2}\delta\gamma_x^\mu
    +
    R^\mu{}_{\nu\rho\sigma}
    k^\nu \delta\gamma_x^\rho k^\sigma
    =
    -
    \delta\Gamma^\mu_{\alpha\beta}
    k^\alpha k^\beta ,
    \label{eq:variation_geodesic_section3}
\end{equation}
where
\begin{equation}
    \delta\Gamma^\mu_{\alpha\beta}
    =
    \frac12 g^{\mu\rho}
    \left(
    \nabla_\alpha\delta g_{\beta\rho}
    +
    \nabla_\beta\delta g_{\alpha\rho}
    -
    \nabla_\rho\delta g_{\alpha\beta}
    \right).
    \label{eq:delta_christoffel_section3}
\end{equation}
At conjugate points, the associated boundary-value problem can fail to be
invertible or can become branch dependent.  Consequently, the metric variation
of \(C_{\rm null}\) is not guaranteed to define a single-valued smooth
functional derivative.

\subsection{Distributional sources and the modeling choice made here}
\label{subsec:distributional_modeling_choice_section3}

The appearance of distributional stress-energy contributions is not, by
itself, a logical inconsistency.  Thin shells, point particles, impulsive waves,
and other distributional sources can be treated in general relativity when the
corresponding junction conditions or weak-field equations are properly defined.
Therefore, the conclusion is not that null-supported memory sources are
mathematically forbidden.

The point is narrower.  The present paper aims to construct a smooth
action-based effective field theory in which the state dependence of the
nonlocal scale is controlled by ordinary covariant fields and in which the
exchange of stress-energy between the memory sector and the history sector is
governed by local Euler--Lagrange equations.  For that purpose, a
line-supported metric variation is not the appropriate object.  It would require
a separate distributional variational framework and would obscure the local
Noether identity used later to prove the conservation of the combined
memory-history stress tensor.

Thus distributional null-memory sources are excluded here as a modeling choice,
not because they are impossible in principle, but because the goal is to obtain
a smooth covariant bulk formulation.  The auxiliary field \(C\) provides such a
formulation.

\subsection{Auxiliary-field replacement}
\label{subsec:auxiliary_replacement_section3}

The replacement adopted in this work is to introduce \(C\) as a scalar field
satisfying a local covariant equation,
\begin{equation}
    (\Box-m_c^2)C
    =
    -\ell_*\mathcal{K}.
    \label{eq:C_auxiliary_preview_section3}
\end{equation}
The source is the same curvature scalar that appears naturally in
Eq.~\eqref{eq:null_memory_integrand_section3}.  The difference is that the
history information is now encoded through the solution of a hyperbolic field
equation rather than by selecting an individual null ray.

When causal boundary data are imposed, the solution may be written in terms of
the retarded Green function of \(\Box-m_c^2\), as discussed in
Sec.~\ref{subsec:boundary_data_section2}.  In that representation, \(C(x)\)
depends on curvature in the causal past of \(x\), but the variational
formulation remains local because \(C\) is treated as an independent field in
the action.

The auxiliary formulation has three advantages for the present purpose.  First,
\(C\) is a scalar under diffeomorphisms.  Second, its equation of motion is
local and covariant.  Third, its stress-energy tensor follows from an ordinary
metric variation of a local action once the conjugate field \(\chi\) is
introduced.  This makes it possible to derive a local on-shell conservation
identity for the combined memory-history sector.

The construction should therefore be understood as a covariant smoothing and
localization of the null-history intuition.  It does not assert that
Eq.~\eqref{eq:C_auxiliary_preview_section3} is the unique possible definition
of curvature memory.  It provides a technically controlled definition suitable
for the state-dependent infinite-derivative action constructed in the next
section.

\section{Covariant Action and Operator Ordering}
\label{sec:covariant_auxiliary_action}

We now define the covariant action used in the rest of the paper.  The
conventions, field content, dimensional assignments, and boundary assumptions
were fixed in Sec.~\ref{sec:geometric_setup}.  The purpose of the present
section is to specify the dynamical functional and, in particular, to remove any
ambiguity in the ordering of the state-dependent nonlocal operator.

The total action is written as
\begin{equation}
    S
    =
    S_{\rm EH}
    +
    S_{\rm mat}
    +
    S_\Delta
    +
    S_C .
    \label{eq:total_action_section4}
\end{equation}
Here \(S_{\rm EH}\) is the Einstein--Hilbert action, \(S_{\rm mat}\) is an
ordinary matter action, \(S_\Delta\) is the memory-sector action, and \(S_C\)
is the auxiliary curvature-history action.

\subsection{Einstein--Hilbert and matter sectors}
\label{subsec:EH_matter_section4}

The gravitational action is
\begin{equation}
    S_{\rm EH}
    =
    \frac{1}{16\pi G_N}
    \int_{\mathcal{M}} d^4x\,\sqrt{-g}\,R .
    \label{eq:EH_action_section4}
\end{equation}
Equivalently, in terms of the reduced Planck mass
\begin{equation}
    M_{\rm Pl}^2
    =
    \frac{1}{8\pi G_N},
\end{equation}
this may be written as
\begin{equation}
    S_{\rm EH}
    =
    \frac{M_{\rm Pl}^2}{2}
    \int_{\mathcal{M}} d^4x\,\sqrt{-g}\,R .
\end{equation}

The matter sector is assumed to be diffeomorphism invariant:
\begin{equation}
    S_{\rm mat}
    =
    \int_{\mathcal{M}} d^4x\,\sqrt{-g}\,
    \mathcal{L}_{\rm mat}(g,\Phi).
    \label{eq:matter_action_general_section4}
\end{equation}
For a minimally coupled scalar field, one may take
\begin{equation}
    \mathcal{L}_{\rm mat}
    =
    -\frac12 g^{\mu\nu}\nabla_\mu\Phi\nabla_\nu\Phi
    -
    U(\Phi),
    \label{eq:matter_scalar_example_section4}
\end{equation}
but the subsequent memory-history construction does not depend on this
particular matter choice.  The only property used below is the
diffeomorphism invariance of \(S_{\rm mat}\).

\subsection{Auxiliary curvature-history sector}
\label{subsec:auxiliary_history_sector_section4}

The curvature-history field \(C\) is introduced as a local scalar field rather
than as an externally prescribed null-congruence integral.  Its equation of
motion is imposed through a conjugate auxiliary field \(\chi\).  The auxiliary
history action is
\begin{equation}
    S_C
    =
    \int_{\mathcal{M}} d^4x\,\sqrt{-g}\,
    \mathcal{L}_C ,
    \label{eq:SC_action_section4}
\end{equation}
with
\begin{equation}
    \mathcal{L}_C
    =
    -
    \nabla_\mu\chi\nabla^\mu C
    -
    m_c^2\chi C
    +
    \ell_*\chi\mathcal{K}.
    \label{eq:LC_section4}
\end{equation}
Here
\begin{equation}
    \mathcal{K}
    =
    R_{\mu\nu\rho\sigma}R^{\mu\nu\rho\sigma}
    \label{eq:K_section4}
\end{equation}
is the Kretschmann scalar, \(m_c\) is an infrared mass scale, and \(\ell_*\) is
a fixed length scale.  With the dimensional assignments of
Sec.~\ref{subsec:dimensional_assignments_section2}, each term in
\(\mathcal{L}_C\) has mass dimension four.

The sign convention in Eq.~\eqref{eq:LC_section4} is chosen so that variation
with respect to \(\chi\) gives
\begin{equation}
    (\Box-m_c^2)C
    =
    -\ell_*\mathcal{K}.
    \label{eq:C_equation_preview_section4}
\end{equation}
Thus \(C\) is sourced by the full tidal-curvature invariant
\(\mathcal{K}\).  The detailed variation is given in
Sec.~\ref{sec:auxiliary_history_equations}.

The local action does not by itself choose a unique solution of
Eq.~\eqref{eq:C_equation_preview_section4}; boundary or initial data must be
specified.  For the history interpretation used in this paper, the causal
choice is the retarded solution discussed in
Sec.~\ref{subsec:boundary_data_section2}.  The local variational identities
derived below are independent of replacing \(C\) by that explicit
Green-function representation.

\subsection{State-dependent nonlocality scale}
\label{subsec:state_dependent_scale_section4}

The nonlocality scale is allowed to depend on the curvature-history field:
\begin{equation}
    M_*^2
    \longrightarrow
    M_{\rm eff}^2(C).
    \label{eq:state_dependent_replacement_section4}
\end{equation}
We assume throughout that
\begin{equation}
    M_{\rm eff}^2(C)>0
    \label{eq:Meff_positive_section4}
\end{equation}
on the field domain considered.  It is useful to write
\begin{equation}
    B(C)
    \equiv
    M_{\rm eff}^{-2}(C).
    \label{eq:B_def_section4}
\end{equation}
Then \(B(C)\) is a smooth scalar multiplication operator with mass dimension
\(-2\).

A representative parametrization is
\begin{equation}
    M_{\rm eff}^2(C)
    =
    M_*^2\,\mathcal{M}\!\left(\frac{C}{C_*}\right),
    \qquad
    \mathcal{M}(z)>0,
    \label{eq:Meff_general_param_section4}
\end{equation}
where \(M_*\) and \(C_*\) have mass dimension one and \(\mathcal{M}\) is a
positive dimensionless function.  For example, one may choose
\begin{equation}
    \mathcal{M}(z)=\frac{1}{1+z}
    \label{eq:Meff_example_section4}
\end{equation}
on a field domain where \(1+z>0\).  No result in this paper depends on this
specific choice; only smoothness and positivity of \(M_{\rm eff}^2(C)\) are
used.

The physical role of \(M_{\rm eff}(C)\) is to let the strength and scale of the
infinite-derivative modification depend on the curvature-history field.  The
field \(C\) is therefore not merely an external regulator.  It is a covariant
state variable whose equation of motion and stress-energy contribution are
included in the action.

\subsection{Operator-ordering convention}
\label{subsec:operator_ordering_section4}

Because \(C\) is spacetime dependent, \(B(C)\) does not generally commute with
the covariant d'Alembertian:
\begin{equation}
    [B(C),\Box]\neq 0 .
    \label{eq:B_box_noncommute_section4}
\end{equation}
Consequently, the expression \(\Box/M_{\rm eff}^2(C)\) is ambiguous unless an
operator ordering is fixed.  In this paper we use the following convention.

Let \(\Box_\Delta\) denote the covariant rough Laplacian acting on symmetric
rank-two tensor fields:
\begin{equation}
    (\Box_\Delta X)_{\mu\nu}
    =
    g^{\rho\sigma}\nabla_\rho\nabla_\sigma X_{\mu\nu}.
    \label{eq:tensor_box_section4}
\end{equation}
We define the dimensionless operator
\begin{equation}
    \mathcal{A}_C
    \equiv
    B(C)\Box_\Delta .
    \label{eq:A_C_def_section4}
\end{equation}
All operator products are read from right to left.  Thus, for any symmetric
rank-two tensor \(X_{\mu\nu}\),
\begin{equation}
    (\mathcal{A}_C X)_{\mu\nu}
    =
    B(C)\,
    (\Box_\Delta X)_{\mu\nu}.
    \label{eq:A_C_action_section4}
\end{equation}
With this convention,
\begin{equation}
    \mathcal{A}_C^2 X
    =
    B(C)\Box_\Delta
    \left[
        B(C)\Box_\Delta X
    \right],
    \label{eq:A_C_squared_section4}
\end{equation}
so derivatives in the outer \(\Box_\Delta\) act both on the inner tensor and on
the intervening factor \(B(C)\).  This ordered composition is the convention
used everywhere below.

The state-dependent form factor is then defined by the analytic functional
calculus
\begin{equation}
    F_C
    \equiv
    F(\Box_\Delta,C)
    =
    \exp(\mathcal{A}_C)
    =
    \sum_{n=0}^{\infty}
    \frac{1}{n!}\mathcal{A}_C^n .
    \label{eq:F_C_def_section4}
\end{equation}
Equivalently, whenever the notation
\begin{equation}
    \exp\!\left(\frac{\Box}{M_{\rm eff}^2(C)}\right)
\end{equation}
is used, it is to be understood as the ordered operator
\begin{equation}
    \exp\!\left[B(C)\Box_\Delta\right].
    \label{eq:ordered_exponential_section4}
\end{equation}
Other orderings, such as \(\Box_\Delta B(C)\), would define different
state-dependent nonlocal models.  They are not used in this work.

The variation of \(F_C\) cannot be computed by treating \(\mathcal{A}_C\) as a
commuting scalar.  The noncommuting variation is handled by Duhamel's formula
in Sec.~\ref{sec:duhamel_variation}.  With the ordering
\(\mathcal{A}_C=B(C)\Box_\Delta\), the \(C\)-variation of the exponent at fixed
metric is
\begin{equation}
    \delta_C\mathcal{A}_C
    =
    B'(C)\,\delta C\,\Box_\Delta ,
    \label{eq:delta_A_C_section4}
\end{equation}
where \(B'(C)=dB/dC\).  This identity fixes the source term generated by the
\(C\)-dependence of the memory-sector action.

\subsection{Memory-sector action}
\label{subsec:memory_sector_section4}

The memory sector is built from the symmetric tensor field
\(\Delta_{\mu\nu}\).  Its action is
\begin{equation}
    S_\Delta
    =
    \int_{\mathcal{M}} d^4x\,\sqrt{-g}\,
    \mathcal{L}_\Delta ,
    \label{eq:SDelta_action_section4}
\end{equation}
where
\begin{equation}
    \mathcal{L}_\Delta
    =
    -
    \frac12
    \Delta_{\mu\nu}
    \left[
        F_C\left(\Box_\Delta\Delta\right)
    \right]^{\mu\nu}
    -
    V(\Delta,C)
    +
    \eta\,G(C)\Delta_{\mu\nu}R^{\mu\nu}.
    \label{eq:LDelta_section4}
\end{equation}
Here \(F_C=\exp(B(C)\Box_\Delta)\) is the ordered form factor defined in
Eq.~\eqref{eq:F_C_def_section4}.  The term
\(\left[F_C(\Box_\Delta\Delta)\right]^{\mu\nu}\) means that
\(\Box_\Delta\) first acts on \(\Delta_{\mu\nu}\), after which the ordered
operator \(F_C\) acts on the resulting symmetric tensor.  All indices are then
raised with the metric.

The potential \(V(\Delta,C)\) is assumed to be a scalar under diffeomorphisms
and differentiable with respect to both \(\Delta_{\mu\nu}\) and \(C\).  A
representative example is
\begin{equation}
    V(\Delta,C)
    =
    \frac{\lambda}{4}
    \left(
        \Delta_{\mu\nu}\Delta^{\mu\nu}
        -
        \Delta_0^2(C)
    \right)^2,
    \qquad
    \lambda>0,
    \label{eq:potential_example_section4}
\end{equation}
where \(\Delta_0(C)\) is a scalar function of \(C\).  The analysis below does
not require this specific form; only the scalar character and differentiability
of \(V\) are used.

The final term in Eq.~\eqref{eq:LDelta_section4},
\begin{equation}
    \eta\,G(C)\Delta_{\mu\nu}R^{\mu\nu},
    \label{eq:Delta_R_coupling_section4}
\end{equation}
couples the memory tensor to Ricci curvature.  The coupling \(\eta\) is taken
dimensionless, while \(G(C)\) is a scalar function chosen so that the product
has mass dimension four.  This term allows the memory field to interact
directly with the local curvature while preserving diffeomorphism invariance.

\subsection{Diffeomorphism invariance}
\label{subsec:diffeomorphism_invariance_section4}

Each term in Eq.~\eqref{eq:total_action_section4} is the integral of a scalar
density.  The Einstein--Hilbert and matter actions are diffeomorphism invariant
by construction.  The auxiliary action \(S_C\) is invariant because
\(\nabla_\mu\chi\nabla^\mu C\), \(\chi C\), and \(\chi\mathcal{K}\) are scalar
quantities.

The memory-sector action is also diffeomorphism invariant.  The tensor
\(\Box_\Delta\Delta_{\mu\nu}\) is defined covariantly, \(B(C)\) is scalar
multiplication, and the ordered form factor \(F_C=\exp(B(C)\Box_\Delta)\) is
built from covariant operations.  The potential \(V(\Delta,C)\) is assumed to
be a scalar, and the nonminimal term \(G(C)\Delta_{\mu\nu}R^{\mu\nu}\) is a full
contraction of tensors with a scalar coefficient.

This diffeomorphism invariance is the structural reason that the metric field
equation is compatible with the contracted Bianchi identity.  In later
sections, the corresponding Noether identity is used to show that the
stress-energy exchange between the memory sector and the auxiliary
curvature-history sector cancels on shell.

\subsection{Summary of the action}
\label{subsec:summary_action_section4}

The complete action used below is therefore
\begin{align}
    S
    &=
    \frac{1}{16\pi G_N}
    \int d^4x\,\sqrt{-g}\,R
    +
    \int d^4x\,\sqrt{-g}\,
    \mathcal{L}_{\rm mat}
    \nonumber \\
    &\quad
    +
    \int d^4x\,\sqrt{-g}
    \left[
        -
        \frac12
        \Delta_{\mu\nu}
        \left[
            F_C\left(\Box_\Delta\Delta\right)
        \right]^{\mu\nu}
        -
        V(\Delta,C)
        +
        \eta\,G(C)\Delta_{\mu\nu}R^{\mu\nu}
    \right]
    \nonumber \\
    &\quad
    +
    \int d^4x\,\sqrt{-g}
    \left[
        -
        \nabla_\mu\chi\nabla^\mu C
        -
        m_c^2\chi C
        +
        \ell_*\chi\mathcal{K}
    \right],
    \label{eq:complete_action_section4}
\end{align}
where
\begin{equation}
    F_C
    =
    \exp\!\left[B(C)\Box_\Delta\right],
    \qquad
    B(C)=M_{\rm eff}^{-2}(C).
    \label{eq:FC_summary_section4}
\end{equation}
This fixes the operator ordering and the field content of the model.  The
Euler--Lagrange equations for \(C\), \(\chi\), and \(\Delta_{\mu\nu}\) are
derived in the following sections.

\section{Auxiliary Curvature-History Equations}
\label{sec:auxiliary_history_equations}

This section derives the Euler--Lagrange equations for the auxiliary
curvature-history fields \(C\) and \(\chi\).  The field \(C\) carries the
curvature-history information, while \(\chi\) is the conjugate auxiliary field
that imposes the history equation at the level of the action.  The derivation is
performed at fixed metric unless stated otherwise.  Metric variation and the
corresponding stress-energy tensors are treated separately in
Sec.~\ref{sec:metric_variation}.

The auxiliary action is
\begin{equation}
    S_C
    =
    \int_{\mathcal{M}} d^4x\,\sqrt{-g}\,
    \left[
        -
        \nabla_\mu\chi\nabla^\mu C
        -
        m_c^2\chi C
        +
        \ell_*\chi\mathcal{K}
    \right],
    \label{eq:SC_section5}
\end{equation}
where
\begin{equation}
    \mathcal{K}
    =
    R_{\mu\nu\rho\sigma}R^{\mu\nu\rho\sigma}.
\end{equation}
The memory-sector action is
\begin{equation}
    S_\Delta
    =
    \int_{\mathcal{M}} d^4x\,\sqrt{-g}
    \left[
        -
        \frac12
        \Delta_{\mu\nu}
        \left[
            F_C\left(\Box_\Delta\Delta\right)
        \right]^{\mu\nu}
        -
        V(\Delta,C)
        +
        \eta G(C)\Delta_{\mu\nu}R^{\mu\nu}
    \right],
    \label{eq:SDelta_section5}
\end{equation}
with
\begin{equation}
    F_C
    =
    \exp(\mathcal{A}_C),
    \qquad
    \mathcal{A}_C
    =
    B(C)\Box_\Delta,
    \qquad
    B(C)=M_{\rm eff}^{-2}(C).
    \label{eq:FC_section5}
\end{equation}
The ordering convention in Eq.~\eqref{eq:FC_section5} is the one fixed in
Sec.~\ref{subsec:operator_ordering_section4}.

\subsection{Variation with respect to \texorpdfstring{\(\chi\)}{chi}}
\label{subsec:variation_chi_section5}

Varying \(S_C\) with respect to \(\chi\), while holding \(g_{\mu\nu}\) and
\(C\) fixed, gives
\begin{align}
    \delta_\chi S_C
    &=
    \int_{\mathcal{M}} d^4x\,\sqrt{-g}\,
    \left[
        -
        \nabla_\mu(\delta\chi)\nabla^\mu C
        -
        m_c^2(\delta\chi)C
        +
        \ell_*(\delta\chi)\mathcal{K}
    \right].
    \label{eq:delta_chi_SC_start_section5}
\end{align}
Using the boundary assumptions stated in
Sec.~\ref{subsec:regularity_assumptions_section2}, the derivative term may be
integrated by parts:
\begin{equation}
    -
    \int_{\mathcal{M}} d^4x\,\sqrt{-g}\,
    \nabla_\mu(\delta\chi)\nabla^\mu C
    =
    \int_{\mathcal{M}} d^4x\,\sqrt{-g}\,
    \delta\chi\,\Box C .
    \label{eq:chi_integration_by_parts_section5}
\end{equation}
Therefore
\begin{equation}
    \delta_\chi S_C
    =
    \int_{\mathcal{M}} d^4x\,\sqrt{-g}\,
    \delta\chi
    \left[
        \Box C
        -
        m_c^2 C
        +
        \ell_*\mathcal{K}
    \right].
    \label{eq:delta_chi_SC_final_section5}
\end{equation}
Stationarity for arbitrary \(\delta\chi\) gives
\begin{equation}
    \Box C
    -
    m_c^2 C
    +
    \ell_*\mathcal{K}
    =
    0.
    \label{eq:C_equation_section5}
\end{equation}
Equivalently,
\begin{equation}
    (\Box-m_c^2)C
    =
    -\ell_*\mathcal{K}.
    \label{eq:C_history_equation_section5}
\end{equation}

This is the local curvature-history equation.  It replaces the formal
single-null-ray memory variable by a scalar field sourced by the Kretschmann
invariant.  A particular solution is selected only after specifying boundary or
initial data.  For the causal history interpretation, the retarded branch
described in Eq.~\eqref{eq:C_retarded_solution_section2} is the relevant one.

\subsection{Variation with respect to \texorpdfstring{\(C\)}{C}}
\label{subsec:variation_C_section5}

Next vary the combined memory-history action \(S_C+S_\Delta\) with respect to
\(C\), while holding \(g_{\mu\nu}\), \(\chi\), and \(\Delta_{\mu\nu}\) fixed.

The variation of the auxiliary sector is
\begin{align}
    \delta_C S_C
    &=
    \int_{\mathcal{M}} d^4x\,\sqrt{-g}\,
    \left[
        -
        \nabla_\mu\chi\nabla^\mu(\delta C)
        -
        m_c^2\chi\,\delta C
    \right].
    \label{eq:delta_C_SC_start_section5}
\end{align}
Integrating the derivative term by parts gives
\begin{equation}
    -
    \int_{\mathcal{M}} d^4x\,\sqrt{-g}\,
    \nabla_\mu\chi\nabla^\mu(\delta C)
    =
    \int_{\mathcal{M}} d^4x\,\sqrt{-g}\,
    \delta C\,\Box\chi .
    \label{eq:C_integration_by_parts_section5}
\end{equation}
Hence
\begin{equation}
    \delta_C S_C
    =
    \int_{\mathcal{M}} d^4x\,\sqrt{-g}\,
    \delta C
    \left[
        \Box\chi
        -
        m_c^2\chi
    \right].
    \label{eq:delta_C_SC_final_section5}
\end{equation}

The memory-sector action depends on \(C\) in three places: the ordered
nonlocal form factor \(F_C\), the potential \(V(\Delta,C)\), and the scalar
coupling \(G(C)\).  We define the memory response source
\(\Sigma_\Delta\) by
\begin{equation}
    \delta_C S_\Delta
    \equiv
    \int_{\mathcal{M}} d^4x\,\sqrt{-g}\,
    \Sigma_\Delta\,\delta C .
    \label{eq:Sigma_definition_section5}
\end{equation}
This definition is local after all integrations by parts and after the
noncommuting variation of \(F_C\) has been expressed in terms of
\(\delta C\).  The detailed Duhamel evaluation of the nonlocal part is given in
Sec.~\ref{sec:duhamel_variation}.

Using Eq.~\eqref{eq:Sigma_definition_section5}, the total variation with
respect to \(C\) is
\begin{equation}
    \delta_C(S_C+S_\Delta)
    =
    \int_{\mathcal{M}} d^4x\,\sqrt{-g}\,
    \delta C
    \left[
        \Box\chi
        -
        m_c^2\chi
        +
        \Sigma_\Delta
    \right].
    \label{eq:total_C_variation_section5}
\end{equation}
Stationarity for arbitrary \(\delta C\) gives
\begin{equation}
    \Box\chi
    -
    m_c^2\chi
    +
    \Sigma_\Delta
    =
    0.
    \label{eq:chi_equation_start_section5}
\end{equation}
Equivalently,
\begin{equation}
    (\Box-m_c^2)\chi
    =
    -\Sigma_\Delta .
    \label{eq:chi_equation_section5}
\end{equation}
Thus \(\chi\) is sourced by the response of the memory sector to changes in the
curvature-history field \(C\).

\subsection{Structure of the source \texorpdfstring{\(\Sigma_\Delta\)}{SigmaDelta}}
\label{subsec:Sigma_structure_section5}

The source \(\Sigma_\Delta\) naturally decomposes into three terms:
\begin{equation}
    \Sigma_\Delta
    =
    \Sigma_{\rm kin}
    -
    \frac{\partial V}{\partial C}
    +
    \eta G'(C)\Delta_{\mu\nu}R^{\mu\nu}.
    \label{eq:Sigma_decomposition_section5}
\end{equation}
Here \(G'(C)=dG/dC\), and \(\partial V/\partial C\) is taken at fixed
\(g_{\mu\nu}\) and fixed \(\Delta_{\mu\nu}\).

The first term \(\Sigma_{\rm kin}\) is produced by the \(C\)-dependence of the
ordered nonlocal kinetic operator:
\begin{equation}
    \delta_C S_{\Delta}^{\rm kin}
    \equiv
    \int_{\mathcal{M}} d^4x\,\sqrt{-g}\,
    \Sigma_{\rm kin}\,\delta C ,
    \label{eq:Sigma_kin_definition_section5}
\end{equation}
where
\begin{equation}
    S_{\Delta}^{\rm kin}
    =
    -
    \frac12
    \int_{\mathcal{M}} d^4x\,\sqrt{-g}\,
    \Delta_{\mu\nu}
    \left[
        F_C\left(\Box_\Delta\Delta\right)
    \right]^{\mu\nu}.
    \label{eq:SDelta_kin_section5}
\end{equation}
At fixed metric, the \(C\)-variation of the exponent is
\begin{equation}
    \delta_C\mathcal{A}_C
    =
    B'(C)\delta C\,\Box_\Delta .
    \label{eq:delta_A_C_section5}
\end{equation}
Since \(\mathcal{A}_C\) and \(\delta_C\mathcal{A}_C\) do not generally commute,
the variation of \(F_C=\exp(\mathcal{A}_C)\) is not
\(F_C\,\delta_C\mathcal{A}_C\).  Instead,
\begin{equation}
    \delta_C F_C
    =
    \int_0^1
    e^{s\mathcal{A}_C}
    \left(
        \delta_C\mathcal{A}_C
    \right)
    e^{(1-s)\mathcal{A}_C}
    ds .
    \label{eq:duhamel_preview_section5}
\end{equation}
This identity is the basis for the explicit expression for
\(\Sigma_{\rm kin}\) derived in Sec.~\ref{sec:duhamel_variation}.

The remaining two terms in Eq.~\eqref{eq:Sigma_decomposition_section5} are
local.  They follow directly from
\begin{equation}
    \delta_C[-V(\Delta,C)]
    =
    -
    \frac{\partial V}{\partial C}\,\delta C ,
    \label{eq:potential_C_variation_section5}
\end{equation}
and
\begin{equation}
    \delta_C
    \left[
        \eta G(C)\Delta_{\mu\nu}R^{\mu\nu}
    \right]
    =
    \eta G'(C)\Delta_{\mu\nu}R^{\mu\nu}\delta C .
    \label{eq:G_C_variation_section5}
\end{equation}

Equation~\eqref{eq:Sigma_decomposition_section5} shows explicitly how the
memory sector acts as a source for the conjugate field \(\chi\).  In turn, this
source is precisely the term that appears in the stress-energy exchange between
the memory sector and the auxiliary curvature-history sector.  The cancellation
of this exchange is established in Sec.~\ref{sec:conservation_identity}.

\subsection{Coupled auxiliary system}
\label{subsec:coupled_auxiliary_system_section5}

Combining the two equations obtained above, the auxiliary curvature-history
system is
\begin{align}
    (\Box-m_c^2)C
    &=
    -\ell_*\mathcal{K},
    \label{eq:C_system_section5}
    \\
    (\Box-m_c^2)\chi
    &=
    -\Sigma_\Delta.
    \label{eq:chi_system_section5}
\end{align}
The first equation states that \(C\) is sourced by tidal curvature.  The second
states that \(\chi\) is sourced by the dependence of the memory sector on
\(C\).  Thus \(C\) records curvature history, while \(\chi\) records the
response of the state-dependent memory sector to variations of that history
field.

The system in Eqs.~\eqref{eq:C_system_section5}--\eqref{eq:chi_system_section5}
is local and covariant.  Its causal interpretation is obtained by supplementing
it with retarded boundary conditions for \(C\), as discussed in
Sec.~\ref{subsec:boundary_data_section2}.  The local variational derivation
itself does not require substituting the explicit retarded Green-function
solution into the action.

\section{Duhamel Variation of the State-Dependent Form Factor}
\label{sec:duhamel_variation}

The memory-sector source \(\Sigma_\Delta\) defined in
Sec.~\ref{sec:auxiliary_history_equations} contains a contribution from the
\(C\)-dependence of the nonlocal form factor \(F_C\).  This contribution must
be treated with care because the curvature-history field \(C(x)\) is
spacetime dependent.  Consequently, multiplication by
\(B(C)=M_{\rm eff}^{-2}(C)\) does not generally commute with the covariant
d'Alembertian.  The variation of \(F_C\) is therefore an operator variation,
not an ordinary scalar chain rule.

The ordering convention fixed in Sec.~\ref{subsec:operator_ordering_section4}
is
\begin{equation}
    F_C
    =
    \exp(\mathcal{A}_C),
    \qquad
    \mathcal{A}_C
    =
    B(C)\Box_\Delta ,
    \label{eq:FC_AC_section6}
\end{equation}
where \(\Box_\Delta\) is the covariant rough Laplacian acting on symmetric
rank-two tensor fields.  All products are ordered from right to left:
\begin{equation}
    (\mathcal{A}_C X)_{\mu\nu}
    =
    B(C)(\Box_\Delta X)_{\mu\nu}.
    \label{eq:AC_action_section6}
\end{equation}
This section derives the corresponding variation of the kinetic part of
\(S_\Delta\) with respect to \(C\).

\subsection{Noncommutativity of the ordered exponent}
\label{subsec:noncommutativity_section6}

Let \(X_{\mu\nu}\) be a smooth symmetric tensor field.  Acting on \(X_{\mu\nu}\),
the commutator between \(B(C)\) and \(\Box_\Delta\) is
\begin{align}
    [\Box_\Delta,B]X_{\mu\nu}
    &=
    \Box_\Delta(BX_{\mu\nu})
    -
    B\Box_\Delta X_{\mu\nu}
    \nonumber \\
    &=
    (\Box B)X_{\mu\nu}
    +
    2(\nabla^\rho B)\nabla_\rho X_{\mu\nu}.
    \label{eq:box_B_commutator_section6}
\end{align}
Thus
\begin{equation}
    [B(C),\Box_\Delta]\neq 0
    \label{eq:B_box_nonzero_section6}
\end{equation}
unless \(B(C)\) is constant or the tensor field lies in a special restricted
class.  It follows that the exponential
\begin{equation}
    \exp[B(C)\Box_\Delta]
\end{equation}
cannot be varied as if \(B(C)\Box_\Delta\) were an ordinary commuting scalar.

At fixed metric and fixed \(\Delta_{\mu\nu}\), the only \(C\)-dependence of
\(\mathcal{A}_C\) is through \(B(C)\).  Therefore
\begin{equation}
    \delta_C\mathcal{A}_C
    =
    B'(C)\,\delta C\,\Box_\Delta ,
    \label{eq:delta_AC_section6}
\end{equation}
where \(B'(C)=dB/dC\), and \(\delta C\) acts by scalar multiplication.  In
general,
\begin{equation}
    [\mathcal{A}_C,\delta_C\mathcal{A}_C]\neq 0.
    \label{eq:A_deltaA_noncommute_section6}
\end{equation}
Hence the replacement
\begin{equation}
    \delta_C e^{\mathcal{A}_C}
    =
    e^{\mathcal{A}_C}\delta_C\mathcal{A}_C
\end{equation}
is not valid.

\subsection{Duhamel formula}
\label{subsec:duhamel_formula_section6}

For a differentiable one-parameter family of operators \(\mathcal{A}(\epsilon)\),
the first variation of the exponential is given by Duhamel's formula
\cite{Bhatia1997,ReedSimon1975}:
\begin{equation}
    \delta e^{\mathcal{A}}
    =
    \int_0^1
    e^{s\mathcal{A}}
    (\delta\mathcal{A})
    e^{(1-s)\mathcal{A}}
    ds .
    \label{eq:duhamel_formula_section6}
\end{equation}
This identity is purely algebraic at the formal level and follows by
differentiating \(e^{\mathcal{A}+\epsilon\delta\mathcal{A}}\) with respect to
\(\epsilon\) at \(\epsilon=0\).  It remains valid for the operator class used
here under the regularity and domain assumptions stated in
Sec.~\ref{subsec:regularity_assumptions_section2}.

Applying Eq.~\eqref{eq:duhamel_formula_section6} to
\(\mathcal{A}_C=B(C)\Box_\Delta\) gives
\begin{equation}
    \delta_C F_C
    =
    \int_0^1
    e^{s\mathcal{A}_C}
    \left[
        B'(C)\delta C\,\Box_\Delta
    \right]
    e^{(1-s)\mathcal{A}_C}
    ds .
    \label{eq:delta_FC_duhamel_section6}
\end{equation}
This is the basic noncommuting variation used in the definition of
\(\Sigma_{\rm kin}\).

When \(\mathcal{A}_C\) and \(\delta_C\mathcal{A}_C\) commute, Duhamel's formula
reduces to the ordinary scalar rule:
\begin{equation}
    \delta e^{\mathcal{A}_C}
    =
    e^{\mathcal{A}_C}\delta\mathcal{A}_C.
\end{equation}
The present construction does not assume this special case.

\subsection{Bilinear pairing and formal adjoint}
\label{subsec:formal_adjoint_section6}

To extract the coefficient of \(\delta C\), introduce the bilinear pairing
\begin{equation}
    \langle X,Y\rangle
    =
    \int_{\mathcal{M}} d^4x\,\sqrt{-g}\,
    X_{\mu\nu}Y^{\mu\nu}
    \label{eq:tensor_pairing_section6}
\end{equation}
for smooth symmetric rank-two tensors.  This pairing is used only as a formal
device for integrations by parts; in Lorentzian signature it is not a positive
inner product.

Under the boundary assumptions of
Sec.~\ref{subsec:regularity_assumptions_section2}, the tensor d'Alembertian is
formally self-adjoint:
\begin{equation}
    \langle X,\Box_\Delta Y\rangle
    =
    \langle \Box_\Delta X,Y\rangle .
    \label{eq:box_self_adjoint_section6}
\end{equation}
Scalar multiplication by the real function \(B(C)\) is also formally
self-adjoint.  Therefore the formal adjoint of
\(\mathcal{A}_C=B(C)\Box_\Delta\) is
\begin{equation}
    \mathcal{A}_C^\dagger
    =
    \Box_\Delta B(C),
    \label{eq:A_adjoint_section6}
\end{equation}
meaning
\begin{equation}
    \langle X,\mathcal{A}_C Y\rangle
    =
    \langle \mathcal{A}_C^\dagger X,Y\rangle.
    \label{eq:A_adjoint_def_section6}
\end{equation}
Explicitly,
\begin{equation}
    (\mathcal{A}_C^\dagger X)_{\mu\nu}
    =
    \Box_\Delta\!\left[B(C)X_{\mu\nu}\right].
    \label{eq:A_adjoint_action_section6}
\end{equation}
Because \(B(C)\) and \(\Box_\Delta\) do not commute,
\begin{equation}
    \mathcal{A}_C^\dagger\neq \mathcal{A}_C
\end{equation}
in general.

The adjoint of the exponential is correspondingly
\begin{equation}
    F_C^\dagger
    =
    e^{\mathcal{A}_C^\dagger}.
    \label{eq:F_adjoint_section6}
\end{equation}
This distinction is important when isolating the local source multiplying
\(\delta C\).

\subsection{Variation of the nonlocal kinetic term}
\label{subsec:variation_nonlocal_kinetic_section6}

The kinetic part of the memory-sector action is
\begin{equation}
    S_\Delta^{\rm kin}
    =
    -
    \frac12
    \int_{\mathcal{M}} d^4x\,\sqrt{-g}\,
    \Delta_{\mu\nu}
    \left[
        F_C(\Box_\Delta\Delta)
    \right]^{\mu\nu}.
    \label{eq:SDelta_kin_section6}
\end{equation}
Define
\begin{equation}
    Y_{\mu\nu}
    \equiv
    (\Box_\Delta\Delta)_{\mu\nu}.
    \label{eq:Y_def_section6}
\end{equation}
Then
\begin{equation}
    S_\Delta^{\rm kin}
    =
    -
    \frac12
    \langle \Delta,F_CY\rangle .
    \label{eq:SDelta_kin_pairing_section6}
\end{equation}

At fixed metric and fixed \(\Delta_{\mu\nu}\), the variation with respect to
\(C\) is
\begin{equation}
    \delta_C S_\Delta^{\rm kin}
    =
    -
    \frac12
    \left\langle
        \Delta,
        (\delta_C F_C)Y
    \right\rangle .
    \label{eq:delta_Skin_start_section6}
\end{equation}
Using Eq.~\eqref{eq:delta_FC_duhamel_section6},
\begin{align}
    \delta_C S_\Delta^{\rm kin}
    &=
    -
    \frac12
    \int_0^1 ds\,
    \left\langle
        \Delta,
        e^{s\mathcal{A}_C}
        \left[
            B'(C)\delta C\,\Box_\Delta
        \right]
        e^{(1-s)\mathcal{A}_C}Y
    \right\rangle .
    \label{eq:delta_Skin_duhamel_section6}
\end{align}
Move \(e^{s\mathcal{A}_C}\) to the first slot using the formal adjoint:
\begin{align}
    \delta_C S_\Delta^{\rm kin}
    &=
    -
    \frac12
    \int_0^1 ds\,
    \left\langle
        e^{s\mathcal{A}_C^\dagger}\Delta,
        B'(C)\delta C\,\Box_\Delta
        e^{(1-s)\mathcal{A}_C}Y
    \right\rangle .
    \label{eq:delta_Skin_adjoint_section6}
\end{align}
Since \(B'(C)\delta C\) is a scalar multiplication operator, this becomes
\begin{align}
    \delta_C S_\Delta^{\rm kin}
    &=
    -
    \frac12
    \int_{\mathcal{M}} d^4x\,\sqrt{-g}\,
    \delta C(x)\,
    B'(C)
    \int_0^1 ds\,
    \left[
        e^{s\mathcal{A}_C^\dagger}\Delta
    \right]_{\mu\nu}
    \left[
        \Box_\Delta
        e^{(1-s)\mathcal{A}_C}Y
    \right]^{\mu\nu}.
    \label{eq:delta_Skin_local_source_section6}
\end{align}
Comparing with the definition
\begin{equation}
    \delta_C S_\Delta^{\rm kin}
    =
    \int_{\mathcal{M}} d^4x\,\sqrt{-g}\,
    \Sigma_{\rm kin}\,\delta C ,
    \label{eq:Sigma_kin_def_section6}
\end{equation}
we obtain
\begin{equation}
    \boxed{
    \Sigma_{\rm kin}
    =
    -
    \frac12
    B'(C)
    \int_0^1 ds\,
    \left[
        e^{s\mathcal{A}_C^\dagger}\Delta
    \right]_{\mu\nu}
    \left[
        \Box_\Delta
        e^{(1-s)\mathcal{A}_C}
        (\Box_\Delta\Delta)
    \right]^{\mu\nu}
    } .
    \label{eq:Sigma_kin_result_section6}
\end{equation}
This is the kinetic contribution to the source \(\Sigma_\Delta\) appearing in
the \(\chi\) equation.

Equation~\eqref{eq:Sigma_kin_result_section6} is written with the operator
ordering fixed in Eq.~\eqref{eq:FC_AC_section6}.  If a different ordering were
chosen for the state-dependent form factor, the adjoint operator and the source
\(\Sigma_{\rm kin}\) would be different.  Thus the expression above is not an
ordering-independent formal identity; it is part of the definition of the
specific model considered in this paper.

\subsection{Complete \texorpdfstring{\(C\)}{C}-source from the memory sector}
\label{subsec:complete_source_section6}

The full \(C\)-variation of \(S_\Delta\) is the sum of the nonlocal kinetic
contribution and the local variations of the potential and curvature coupling:
\begin{equation}
    \delta_C S_\Delta
    =
    \int_{\mathcal{M}} d^4x\,\sqrt{-g}\,
    \Sigma_\Delta\,\delta C .
    \label{eq:delta_SDelta_full_section6}
\end{equation}
Using Eq.~\eqref{eq:Sigma_kin_result_section6}, one finds
\begin{equation}
    \boxed{
    \Sigma_\Delta
    =
    \Sigma_{\rm kin}
    -
    \frac{\partial V}{\partial C}
    +
    \eta G'(C)\Delta_{\mu\nu}R^{\mu\nu}
    } .
    \label{eq:Sigma_full_result_section6}
\end{equation}
The derivative \(\partial V/\partial C\) is taken at fixed
\(g_{\mu\nu}\) and fixed \(\Delta_{\mu\nu}\), while \(G'(C)=dG/dC\).

Substitution into the auxiliary equation obtained in
Sec.~\ref{sec:auxiliary_history_equations} gives
\begin{equation}
    (\Box-m_c^2)\chi
    =
    -
    \Sigma_\Delta .
    \label{eq:chi_equation_with_source_section6}
\end{equation}
Thus the response of the nonlocal memory sector to the state-dependent scale
acts as a source for the conjugate history field \(\chi\).

\subsection{Constant-\texorpdfstring{\(C\)}{C} limit}
\label{subsec:constant_C_limit_section6}

It is useful to record the simplification that occurs when \(C\) is constant on
the region of interest.  In that case \(B(C)\) is constant and
\begin{equation}
    [B(C),\Box_\Delta]=0.
\end{equation}
The exponent becomes
\begin{equation}
    \mathcal{A}_C
    =
    B(C)\Box_\Delta,
    \qquad
    \mathcal{A}_C^\dagger=\mathcal{A}_C,
\end{equation}
and the form factor reduces to the standard fixed-scale infinite-derivative
operator
\begin{equation}
    F_C
    =
    \exp\!\left[B(C)\Box_\Delta\right].
\end{equation}
If, in addition, \(\delta C\) is treated as a constant variation, then
\(\delta_C\mathcal{A}_C\) commutes with \(\mathcal{A}_C\), and Duhamel's
formula reduces to the ordinary expression
\begin{equation}
    \delta_C F_C
    =
    F_C\,B'(C)\delta C\,\Box_\Delta .
\end{equation}
The general formula in Eq.~\eqref{eq:Sigma_kin_result_section6} is therefore
the covariant spacetime-dependent extension of the familiar fixed-scale
variation.

\subsection{Role in the conservation identity}
\label{subsec:duhamel_role_conservation_section6}

The source \(\Sigma_\Delta\) defined in
Eq.~\eqref{eq:Sigma_full_result_section6} has two roles.  First, it appears as
the source in the auxiliary equation for \(\chi\).  Second, it measures the
failure of the memory-sector stress tensor to be separately conserved when
\(C\) is spacetime dependent.  On the memory-field equation of motion, the
Noether identity for \(S_\Delta\) gives a term proportional to
\(-\Sigma_\Delta\nabla_\nu C\).  The auxiliary history sector gives the
opposite term, \(+\Sigma_\Delta\nabla_\nu C\).  The two contributions cancel in
the combined memory-history stress tensor.

Thus the Duhamel variation is not a purely technical refinement.  It fixes the
precise source that is required for the later on-shell conservation identity to
be correct when the nonlocality scale depends on the dynamical field \(C(x)\).

\section{Memory-Field Equation}
\label{sec:memory_field_equation}

This section derives the Euler--Lagrange equation for the symmetric memory
tensor \(\Delta_{\mu\nu}\).  The variation is performed at fixed
\(g_{\mu\nu}\), fixed \(C\), and fixed \(\chi\).  Therefore the state-dependent
form factor \(F_C\) is treated as a fixed linear operator in this section.  Its
variation with respect to \(C\) was derived in
Sec.~\ref{sec:duhamel_variation}.

The memory-sector action is
\begin{equation}
    S_\Delta
    =
    \int_{\mathcal{M}} d^4x\,\sqrt{-g}
    \left[
        -
        \frac12
        \Delta_{\mu\nu}
        \left[
            F_C(\Box_\Delta\Delta)
        \right]^{\mu\nu}
        -
        V(\Delta,C)
        +
        \eta G(C)\Delta_{\mu\nu}R^{\mu\nu}
    \right],
    \label{eq:SDelta_section7}
\end{equation}
where
\begin{equation}
    F_C
    =
    \exp(\mathcal{A}_C),
    \qquad
    \mathcal{A}_C=B(C)\Box_\Delta,
    \qquad
    B(C)=M_{\rm eff}^{-2}(C).
    \label{eq:FC_section7}
\end{equation}
The ordering convention is the one fixed in
Sec.~\ref{subsec:operator_ordering_section4}.  The field
\(\Delta_{\mu\nu}\) is symmetric, and variations are restricted to symmetric
variations,
\begin{equation}
    \delta\Delta_{\mu\nu}
    =
    \delta\Delta_{\nu\mu}.
    \label{eq:symmetric_delta_variation_section7}
\end{equation}

\subsection{Variation of the nonlocal kinetic term}
\label{subsec:variation_memory_kinetic_section7}

The kinetic part of the memory action is
\begin{equation}
    S_\Delta^{\rm kin}
    =
    -
    \frac12
    \int_{\mathcal{M}} d^4x\,\sqrt{-g}\,
    \Delta_{\mu\nu}
    \left[
        F_C(\Box_\Delta\Delta)
    \right]^{\mu\nu}.
    \label{eq:SDelta_kin_section7}
\end{equation}
Using the bilinear pairing
\begin{equation}
    \langle X,Y\rangle
    =
    \int_{\mathcal{M}} d^4x\,\sqrt{-g}\,
    X_{\mu\nu}Y^{\mu\nu},
    \label{eq:pairing_section7}
\end{equation}
this term can be written as
\begin{equation}
    S_\Delta^{\rm kin}
    =
    -
    \frac12
    \left\langle
        \Delta,
        F_C\Box_\Delta\Delta
    \right\rangle .
    \label{eq:SDelta_kin_pairing_section7}
\end{equation}

At fixed \(C\) and fixed metric, the operator \(F_C\) does not vary under
\(\delta\Delta_{\mu\nu}\).  Therefore
\begin{align}
    \delta_\Delta S_\Delta^{\rm kin}
    &=
    -
    \frac12
    \left\langle
        \delta\Delta,
        F_C\Box_\Delta\Delta
    \right\rangle
    -
    \frac12
    \left\langle
        \Delta,
        F_C\Box_\Delta\delta\Delta
    \right\rangle .
    \label{eq:delta_Skin_start_section7}
\end{align}
The second term must be treated using the formal adjoint of the ordered
operator.  As shown in Sec.~\ref{subsec:formal_adjoint_section6},
\begin{equation}
    F_C^\dagger
    =
    e^{\mathcal{A}_C^\dagger},
    \qquad
    \mathcal{A}_C^\dagger
    =
    \Box_\Delta B(C).
    \label{eq:FC_adjoint_section7}
\end{equation}
Thus
\begin{equation}
    \left\langle
        \Delta,
        F_C\Box_\Delta\delta\Delta
    \right\rangle
    =
    \left\langle
        F_C^\dagger\Delta,
        \Box_\Delta\delta\Delta
    \right\rangle .
    \label{eq:move_F_adjoint_section7}
\end{equation}
Using the formal self-adjointness of \(\Box_\Delta\) under the boundary
assumptions of Sec.~\ref{subsec:regularity_assumptions_section2},
\begin{equation}
    \left\langle
        F_C^\dagger\Delta,
        \Box_\Delta\delta\Delta
    \right\rangle
    =
    \left\langle
        \Box_\Delta F_C^\dagger\Delta,
        \delta\Delta
    \right\rangle .
    \label{eq:move_box_section7}
\end{equation}
Hence
\begin{equation}
    \delta_\Delta S_\Delta^{\rm kin}
    =
    -
    \frac12
    \int_{\mathcal{M}} d^4x\,\sqrt{-g}\,
    \delta\Delta_{\mu\nu}
    \left\{
        \left[
            F_C(\Box_\Delta\Delta)
        \right]^{\mu\nu}
        +
        \left[
            \Box_\Delta(F_C^\dagger\Delta)
        \right]^{\mu\nu}
    \right\}.
    \label{eq:delta_Skin_final_section7}
\end{equation}
Because \(F_C\) is not generally self-adjoint for spacetime-dependent \(C\),
both terms in Eq.~\eqref{eq:delta_Skin_final_section7} must be retained.

\subsection{Variation of the local memory terms}
\label{subsec:variation_local_memory_terms_section7}

The local part of the memory action is
\begin{equation}
    S_\Delta^{\rm loc}
    =
    \int_{\mathcal{M}} d^4x\,\sqrt{-g}
    \left[
        -
        V(\Delta,C)
        +
        \eta G(C)\Delta_{\mu\nu}R^{\mu\nu}
    \right].
    \label{eq:SDelta_local_section7}
\end{equation}
At fixed metric and fixed \(C\), its variation is
\begin{equation}
    \delta_\Delta S_\Delta^{\rm loc}
    =
    \int_{\mathcal{M}} d^4x\,\sqrt{-g}\,
    \left[
        -
        \mathcal{V}^{\mu\nu}
        +
        \eta G(C)R^{\mu\nu}
    \right]
    \delta\Delta_{\mu\nu},
    \label{eq:delta_Slocal_section7}
\end{equation}
where
\begin{equation}
    \mathcal{V}^{\mu\nu}
    \equiv
    \left.
    \frac{\partial V}{\partial\Delta_{\mu\nu}}
    \right|_{g,C}
    \label{eq:V_derivative_section7}
\end{equation}
is the derivative of the scalar potential with respect to the symmetric tensor
\(\Delta_{\mu\nu}\), holding the metric and \(C\) fixed.  Since
\(\delta\Delta_{\mu\nu}\) is symmetric, only the symmetric part of
\(\mathcal{V}^{\mu\nu}\) contributes.  For potentials constructed from scalar
contractions of \(\Delta_{\mu\nu}\), \(\mathcal{V}^{\mu\nu}\) is symmetric
automatically.

For the representative potential
\begin{equation}
    V(\Delta,C)
    =
    \frac{\lambda}{4}
    \left(
        \Delta_{\alpha\beta}\Delta^{\alpha\beta}
        -
        \Delta_0^2(C)
    \right)^2,
    \label{eq:potential_example_section7}
\end{equation}
one obtains
\begin{equation}
    \mathcal{V}^{\mu\nu}
    =
    \lambda
    \left(
        \Delta_{\alpha\beta}\Delta^{\alpha\beta}
        -
        \Delta_0^2(C)
    \right)
    \Delta^{\mu\nu}.
    \label{eq:V_derivative_example_section7}
\end{equation}
The general derivation below does not require this special form.

\subsection{Euler--Lagrange equation for the memory tensor}
\label{subsec:memory_eom_section7}

Combining Eqs.~\eqref{eq:delta_Skin_final_section7} and
\eqref{eq:delta_Slocal_section7}, the full variation of \(S_\Delta\) with
respect to \(\Delta_{\mu\nu}\) is
\begin{align}
    \delta_\Delta S_\Delta
    =
    \int_{\mathcal{M}} d^4x\,\sqrt{-g}\,
    \delta\Delta_{\mu\nu}
    \Bigg[
        &
        -
        \frac12
        \left[
            F_C(\Box_\Delta\Delta)
        \right]^{\mu\nu}
        -
        \frac12
        \left[
            \Box_\Delta(F_C^\dagger\Delta)
        \right]^{\mu\nu}
        \nonumber \\
        &
        -
        \mathcal{V}^{\mu\nu}
        +
        \eta G(C)R^{\mu\nu}
    \Bigg].
    \label{eq:delta_SDelta_full_section7}
\end{align}
Stationarity for arbitrary symmetric \(\delta\Delta_{\mu\nu}\) gives
\begin{equation}
    -
    \frac12
    \left[
        F_C(\Box_\Delta\Delta)
    \right]^{\mu\nu}
    -
    \frac12
    \left[
        \Box_\Delta(F_C^\dagger\Delta)
    \right]^{\mu\nu}
    -
    \mathcal{V}^{\mu\nu}
    +
    \eta G(C)R^{\mu\nu}
    =
    0.
    \label{eq:memory_eom_raw_section7}
\end{equation}
Equivalently,
\begin{equation}
    \boxed{
    \left[
        F_C(\Box_\Delta\Delta)
    \right]^{\mu\nu}
    +
    \left[
        \Box_\Delta(F_C^\dagger\Delta)
    \right]^{\mu\nu}
    +
    2\mathcal{V}^{\mu\nu}
    -
    2\eta G(C)R^{\mu\nu}
    =
    0
    } .
    \label{eq:memory_eom_section7}
\end{equation}

It is useful to define the memory Euler tensor
\begin{equation}
    \mathcal{E}_\Delta^{\mu\nu}
    \equiv
    -
    \frac12
    \left[
        F_C(\Box_\Delta\Delta)
    \right]^{\mu\nu}
    -
    \frac12
    \left[
        \Box_\Delta(F_C^\dagger\Delta)
    \right]^{\mu\nu}
    -
    \mathcal{V}^{\mu\nu}
    +
    \eta G(C)R^{\mu\nu}.
    \label{eq:E_delta_def_section7}
\end{equation}
Then the memory-field equation is simply
\begin{equation}
    \mathcal{E}_\Delta^{\mu\nu}=0.
    \label{eq:E_delta_zero_section7}
\end{equation}

\subsection{Constant-history limit}
\label{subsec:constant_history_limit_section7}

When \(C\) is constant on the region of interest, \(B(C)\) is constant and
\begin{equation}
    [B(C),\Box_\Delta]=0.
\end{equation}
The ordered operator becomes formally self-adjoint:
\begin{equation}
    \mathcal{A}_C^\dagger=\mathcal{A}_C,
    \qquad
    F_C^\dagger=F_C.
\end{equation}
Moreover \(F_C\) commutes with \(\Box_\Delta\).  In this limit,
Eq.~\eqref{eq:memory_eom_section7} reduces to
\begin{equation}
    2
    \left[
        F_C(\Box_\Delta\Delta)
    \right]^{\mu\nu}
    +
    2\mathcal{V}^{\mu\nu}
    -
    2\eta G(C)R^{\mu\nu}
    =
    0,
\end{equation}
or
\begin{equation}
    \left[
        F_C(\Box_\Delta\Delta)
    \right]^{\mu\nu}
    +
    \mathcal{V}^{\mu\nu}
    -
    \eta G(C)R^{\mu\nu}
    =
    0.
    \label{eq:memory_eom_constant_C_section7}
\end{equation}
Thus the general equation
\eqref{eq:memory_eom_section7} reduces to the expected fixed-scale
infinite-derivative form when the curvature-history field is constant.

\subsection{Interpretation}
\label{subsec:memory_eom_interpretation_section7}

Equation~\eqref{eq:memory_eom_section7} shows how the memory tensor is driven by
three structures.  The first two terms contain the ordered nonlocal kinetic
operator and its formal adjoint.  Both are required because the
state-dependent form factor is not self-adjoint for nonconstant \(C\) under the
chosen ordering.  The third term, \(2\mathcal{V}^{\mu\nu}\), encodes the local
self-interaction of the memory field.  The final term,
\(-2\eta G(C)R^{\mu\nu}\), couples the memory tensor to local Ricci curvature
with a strength controlled by the curvature-history scalar.

The memory equation is used in the Noether identity for the memory-sector
stress tensor.  In particular, when
\(\mathcal{E}_\Delta^{\mu\nu}=0\), the remaining nonconservation of
\(T_{\mu\nu}^{(\Delta)}\) is entirely due to the explicit spacetime dependence
introduced through \(C\).  That term is proportional to
\(\Sigma_\Delta\nabla_\nu C\) and is cancelled by the corresponding term from
the auxiliary curvature-history sector in the combined conservation law.

\section{Metric Variation and Stress-Energy Tensors}
\label{sec:metric_variation}

We now vary the action with respect to the metric and define the corresponding
stress-energy tensors.  Throughout this section the fields
\(\Phi\), \(\Delta_{\mu\nu}\), \(C\), and \(\chi\) are treated as independent
fields under metric variation.  In particular, the retarded Green-function
representation of \(C\) is not substituted into the action before variation.
This is essential: the metric variation is performed in the local auxiliary
field formulation, not in a nonlocal reduced formulation in which \(C\) has
already been eliminated.

The total action is
\begin{equation}
    S
    =
    S_{\rm EH}
    +
    S_{\rm mat}
    +
    S_\Delta
    +
    S_C .
    \label{eq:total_action_section8}
\end{equation}
The metric variation is written in the form
\begin{equation}
    \delta_g S
    =
    \delta_g S_{\rm EH}
    +
    \delta_g S_{\rm mat}
    +
    \delta_g S_\Delta
    +
    \delta_g S_C .
    \label{eq:metric_variation_total_section8}
\end{equation}
Boundary contributions from the Einstein--Hilbert action and from integrations
by parts are assumed to be cancelled by the appropriate boundary terms or to
vanish under the boundary conditions stated in
Sec.~\ref{subsec:regularity_assumptions_section2}.

\subsection{Variational definitions}
\label{subsec:stress_tensor_definitions_section8}

The matter stress-energy tensor is defined by
\begin{equation}
    T_{\mu\nu}^{\rm mat}
    =
    -
    \frac{2}{\sqrt{-g}}
    \frac{\delta S_{\rm mat}}{\delta g^{\mu\nu}} .
    \label{eq:Tmat_def_section8}
\end{equation}
Similarly, the memory-sector and auxiliary curvature-history stress-energy
tensors are
\begin{equation}
    T_{\mu\nu}^{(\Delta)}
    =
    -
    \frac{2}{\sqrt{-g}}
    \frac{\delta S_\Delta}{\delta g^{\mu\nu}},
    \label{eq:TDelta_def_section8}
\end{equation}
and
\begin{equation}
    T_{\mu\nu}^{(C)}
    =
    -
    \frac{2}{\sqrt{-g}}
    \frac{\delta S_C}{\delta g^{\mu\nu}}.
    \label{eq:TC_def_section8}
\end{equation}
With these definitions,
\begin{equation}
    \delta_g S_{\rm mat}
    =
    -
    \frac12
    \int_{\mathcal{M}}d^4x\,\sqrt{-g}\,
    T_{\mu\nu}^{\rm mat}\delta g^{\mu\nu},
    \label{eq:delta_Smat_stress_section8}
\end{equation}
and analogously for \(S_\Delta\) and \(S_C\).

The Einstein--Hilbert variation gives
\begin{equation}
    \delta_g S_{\rm EH}
    =
    \frac{1}{16\pi G_N}
    \int_{\mathcal{M}}d^4x\,\sqrt{-g}\,
    G_{\mu\nu}\delta g^{\mu\nu},
    \label{eq:delta_EH_section8}
\end{equation}
up to the boundary terms already mentioned.

\subsection{Auxiliary curvature-history stress tensor}
\label{subsec:auxiliary_stress_tensor_section8}

The auxiliary curvature-history action is
\begin{equation}
    S_C
    =
    \int_{\mathcal{M}} d^4x\,\sqrt{-g}
    \left[
        -
        \nabla_\rho\chi\nabla^\rho C
        -
        m_c^2\chi C
        +
        \ell_*\chi\mathcal{K}
    \right].
    \label{eq:SC_section8}
\end{equation}
It is useful to split the corresponding stress tensor into two parts:
\begin{equation}
    T_{\mu\nu}^{(C)}
    =
    T_{\mu\nu}^{(C,0)}
    +
    \ell_*\,\Theta_{\mu\nu}[\chi],
    \label{eq:TC_split_section8}
\end{equation}
where \(T_{\mu\nu}^{(C,0)}\) comes from the derivative and mass terms,
whereas \(\Theta_{\mu\nu}[\chi]\) comes from the curvature coupling
\(\chi\mathcal{K}\).

The elementary part follows directly from variation of
\begin{equation}
    \mathcal{L}_{C,0}
    =
    -
    \nabla_\rho\chi\nabla^\rho C
    -
    m_c^2\chi C .
    \label{eq:LC0_section8}
\end{equation}
Since \(C\) and \(\chi\) are scalars,
\(\nabla_\mu C=\partial_\mu C\) and
\(\nabla_\mu\chi=\partial_\mu\chi\), so the metric enters this part only
through the contraction and the volume element.  One obtains
\begin{equation}
    T_{\mu\nu}^{(C,0)}
    =
    \nabla_\mu\chi\nabla_\nu C
    +
    \nabla_\nu\chi\nabla_\mu C
    +
    g_{\mu\nu}
    \left[
        -
        \nabla_\rho\chi\nabla^\rho C
        -
        m_c^2\chi C
    \right].
    \label{eq:TC0_section8}
\end{equation}
Equivalently,
\begin{equation}
    T_{\mu\nu}^{(C,0)}
    =
    2\nabla_{(\mu}\chi\nabla_{\nu)}C
    -
    g_{\mu\nu}\nabla_\rho\chi\nabla^\rho C
    -
    g_{\mu\nu}m_c^2\chi C .
    \label{eq:TC0_symmetric_section8}
\end{equation}

The Kretschmann-history contribution is defined by
\begin{equation}
    \Theta_{\mu\nu}[\chi]
    \equiv
    -
    \frac{2}{\sqrt{-g}}
    \frac{\delta}{\delta g^{\mu\nu}}
    \int_{\mathcal{M}}d^4x\,\sqrt{-g}\,
    \chi\mathcal{K}.
    \label{eq:Theta_def_section8}
\end{equation}
For a smooth scalar multiplier \(f\), define more generally
\begin{equation}
    \Theta_{\mu\nu}[f]
    \equiv
    -
    \frac{2}{\sqrt{-g}}
    \frac{\delta}{\delta g^{\mu\nu}}
    \int_{\mathcal{M}}d^4x\,\sqrt{-g}\,
    f R_{\alpha\beta\rho\sigma}R^{\alpha\beta\rho\sigma}.
    \label{eq:Theta_general_def_section8}
\end{equation}
With the curvature convention of Eq.~\eqref{eq:riemann_convention}, and after
discarding boundary terms, this tensor can be written as
\begin{equation}
    \Theta_{\mu\nu}[f]
    =
    g_{\mu\nu}f\mathcal{K}
    -
    4f R_{\mu\alpha\beta\gamma}
    R_\nu{}^{\alpha\beta\gamma}
    -
    8\nabla^\alpha\nabla^\beta
    \left(
        f R_{\mu\alpha\nu\beta}
    \right).
    \label{eq:Theta_general_section8}
\end{equation}
The auxiliary curvature-history contribution in
Eq.~\eqref{eq:TC_split_section8} is obtained by setting
\begin{equation}
    f=\chi.
    \label{eq:f_chi_section8}
\end{equation}
Thus
\begin{equation}
    \Theta_{\mu\nu}[\chi]
    =
    g_{\mu\nu}\chi\mathcal{K}
    -
    4\chi R_{\mu\alpha\beta\gamma}
    R_\nu{}^{\alpha\beta\gamma}
    -
    8\nabla^\alpha\nabla^\beta
    \left(
        \chi R_{\mu\alpha\nu\beta}
    \right).
    \label{eq:Theta_chi_section8}
\end{equation}

Equations~\eqref{eq:TC_split_section8},
\eqref{eq:TC0_symmetric_section8}, and
\eqref{eq:Theta_chi_section8} give the auxiliary stress tensor:
\begin{align}
    T_{\mu\nu}^{(C)}
    &=
    2\nabla_{(\mu}\chi\nabla_{\nu)}C
    -
    g_{\mu\nu}\nabla_\rho\chi\nabla^\rho C
    -
    g_{\mu\nu}m_c^2\chi C
    \nonumber \\
    &\quad
    +
    \ell_*
    \left[
        g_{\mu\nu}\chi\mathcal{K}
        -
        4\chi R_{\mu\alpha\beta\gamma}
        R_\nu{}^{\alpha\beta\gamma}
        -
        8\nabla^\alpha\nabla^\beta
        \left(
            \chi R_{\mu\alpha\nu\beta}
        \right)
    \right].
    \label{eq:TC_full_section8}
\end{align}
The last term contains the higher-curvature metric response generated by the
Kretschmann source in the curvature-history equation.

\subsection{Memory-sector stress tensor}
\label{subsec:memory_stress_tensor_section8}

The memory-sector stress tensor is defined variationally by
Eq.~\eqref{eq:TDelta_def_section8}.  It receives contributions from the
nonlocal kinetic term, the potential, and the Ricci coupling:
\begin{equation}
    T_{\mu\nu}^{(\Delta)}
    =
    T_{\mu\nu}^{(\Delta,{\rm kin})}
    +
    T_{\mu\nu}^{(\Delta,V)}
    +
    T_{\mu\nu}^{(\Delta,R)} .
    \label{eq:TDelta_split_section8}
\end{equation}

The kinetic contribution is
\begin{equation}
    T_{\mu\nu}^{(\Delta,{\rm kin})}
    =
    -
    \frac{2}{\sqrt{-g}}
    \frac{\delta S_\Delta^{\rm kin}}{\delta g^{\mu\nu}},
    \label{eq:TDelta_kin_def_section8}
\end{equation}
where
\begin{equation}
    S_\Delta^{\rm kin}
    =
    -
    \frac12
    \int_{\mathcal{M}}d^4x\,\sqrt{-g}\,
    \Delta_{\alpha\beta}
    \left[
        F_C(\Box_\Delta\Delta)
    \right]^{\alpha\beta}.
    \label{eq:SDelta_kin_section8}
\end{equation}
The metric variation of this term includes the variation of the volume element,
the raising of tensor indices, the connection inside \(\Box_\Delta\), and the
metric dependence of the ordered operator
\begin{equation}
    F_C
    =
    \exp[B(C)\Box_\Delta].
\end{equation}
At fixed \(C\), the scalar multiplier \(B(C)\) is not varied directly, but
\(\Box_\Delta\) and the tensor contractions are varied.  The compact
variational definition in Eq.~\eqref{eq:TDelta_kin_def_section8} is used in the
main text because the fully expanded expression is lengthy and does not alter
the conservation argument, which follows from diffeomorphism invariance.

The potential contribution is
\begin{equation}
    T_{\mu\nu}^{(\Delta,V)}
    =
    -
    \frac{2}{\sqrt{-g}}
    \frac{\delta}{\delta g^{\mu\nu}}
    \left[
        -
        \int_{\mathcal{M}}d^4x\,\sqrt{-g}\,
        V(\Delta,C)
    \right].
    \label{eq:TDelta_V_def_section8}
\end{equation}
Equivalently,
\begin{equation}
    T_{\mu\nu}^{(\Delta,V)}
    =
    -g_{\mu\nu}V
    +
    2
    \left.
    \frac{\partial V}{\partial g^{\mu\nu}}
    \right|_{\Delta,C},
    \label{eq:TDelta_V_section8}
\end{equation}
where the partial derivative accounts for the metric dependence of scalar
contractions inside \(V\).  For a potential written purely in terms of
\(\Delta_{\alpha\beta}\Delta^{\alpha\beta}\), this term includes the variation
of the raised tensor \(\Delta^{\alpha\beta}\).

The Ricci-coupling contribution is
\begin{equation}
    T_{\mu\nu}^{(\Delta,R)}
    =
    -
    \frac{2}{\sqrt{-g}}
    \frac{\delta}{\delta g^{\mu\nu}}
    \int_{\mathcal{M}}d^4x\,\sqrt{-g}\,
    \eta G(C)\Delta_{\alpha\beta}R^{\alpha\beta}.
    \label{eq:TDelta_R_def_section8}
\end{equation}
This term contains the variation of the volume element, the inverse metrics
used in \(R^{\alpha\beta}\), and the variation of the Ricci tensor.  Since
\(C\) and \(\Delta_{\alpha\beta}\) are independent fields under metric
variation, \(G(C)\) and \(\Delta_{\alpha\beta}\) are held fixed before index
raising.  The explicit expanded form is not required for the local Noether
identity; its variational definition is sufficient.

Thus \(T_{\mu\nu}^{(\Delta)}\) is the unique symmetric stress tensor obtained
from metric variation of the diffeomorphism-invariant memory action.  The fact
that \(F_C\) is an ordered nonlocal operator affects the detailed form of
\(T_{\mu\nu}^{(\Delta,{\rm kin})}\), but not the variational identity used in
Sec.~\ref{sec:conservation_identity}.

\subsection{Metric field equation}
\label{subsec:metric_field_equation_section8}

Combining the metric variations, one obtains
\begin{align}
    \delta_g S
    =
    \frac12
    \int_{\mathcal{M}}d^4x\,\sqrt{-g}
    \left[
        \frac{1}{8\pi G_N}G_{\mu\nu}
        -
        T_{\mu\nu}^{\rm mat}
        -
        T_{\mu\nu}^{(\Delta)}
        -
        T_{\mu\nu}^{(C)}
    \right]
    \delta g^{\mu\nu}.
    \label{eq:delta_total_metric_section8}
\end{align}
Stationarity with respect to arbitrary metric variations gives
\begin{equation}
    \frac{1}{8\pi G_N}G_{\mu\nu}
    =
    T_{\mu\nu}^{\rm mat}
    +
    T_{\mu\nu}^{(\Delta)}
    +
    T_{\mu\nu}^{(C)}.
    \label{eq:metric_equation_unscaled_section8}
\end{equation}
Equivalently,
\begin{equation}
    \boxed{
    G_{\mu\nu}
    =
    8\pi G_N
    \left(
        T_{\mu\nu}^{\rm mat}
        +
        T_{\mu\nu}^{(\Delta)}
        +
        T_{\mu\nu}^{(C)}
    \right)
    } .
    \label{eq:modified_einstein_equation_section8}
\end{equation}

Because the left-hand side satisfies the contracted Bianchi identity,
\begin{equation}
    \nabla^\mu G_{\mu\nu}=0,
    \label{eq:bianchi_section8}
\end{equation}
the right-hand side of Eq.~\eqref{eq:modified_einstein_equation_section8} must
be covariantly conserved on shell.  If the matter sector is minimally coupled
and satisfies its own equation of motion, then
\begin{equation}
    \nabla^\mu T_{\mu\nu}^{\rm mat}=0.
    \label{eq:matter_conservation_section8}
\end{equation}
The remaining consistency condition is therefore
\begin{equation}
    \nabla^\mu
    \left(
        T_{\mu\nu}^{(\Delta)}
        +
        T_{\mu\nu}^{(C)}
    \right)
    =
    0.
    \label{eq:memory_history_conservation_needed_section8}
\end{equation}
This identity is not imposed by hand.  It follows from the diffeomorphism
invariance of \(S_\Delta+S_C\) and from the equations of motion for
\(\Delta_{\mu\nu}\), \(C\), and \(\chi\), as shown in the next section.

\section{On-Shell Covariant Conservation}
\label{sec:conservation_identity}

The metric equation derived in Sec.~\ref{sec:metric_variation} is compatible
with the contracted Bianchi identity only if the total stress-energy tensor is
covariantly conserved on shell.  This section proves the required conservation
law for the combined memory-history sector:
\begin{equation}
    \nabla^\mu
    \left(
        T_{\mu\nu}^{(\Delta)}
        +
        T_{\mu\nu}^{(C)}
    \right)
    =
    0 .
    \label{eq:combined_conservation_goal_section9}
\end{equation}
The proof is a Noether identity following from diffeomorphism invariance of
\(S_\Delta+S_C\).  It does not require substituting the retarded Green-function
solution for \(C\) into the action.  The fields \(C\), \(\chi\), and
\(\Delta_{\mu\nu}\) are treated as local independent fields, and the identity is
then evaluated on their equations of motion.

Throughout this section we use the active infinitesimal diffeomorphism
variation generated by a compactly supported vector field \(\xi^\mu\):
\begin{equation}
    \delta_\xi = \mathcal{L}_\xi ,
    \label{eq:lie_variation_convention_section9}
\end{equation}
where \(\mathcal{L}_\xi\) is the Lie derivative.  Therefore
\begin{equation}
    \delta_\xi g^{\mu\nu}
    =
    \mathcal{L}_\xi g^{\mu\nu}
    =
    -
    2\nabla^{(\mu}\xi^{\nu)},
    \label{eq:metric_lie_variation_section9}
\end{equation}
and, for a scalar \(f\),
\begin{equation}
    \delta_\xi f
    =
    \mathcal{L}_\xi f
    =
    \xi^\rho\nabla_\rho f .
    \label{eq:scalar_lie_variation_section9}
\end{equation}
For the symmetric tensor \(\Delta_{\mu\nu}\),
\begin{equation}
    \delta_\xi\Delta_{\mu\nu}
    =
    \mathcal{L}_\xi\Delta_{\mu\nu}
    =
    \xi^\rho\nabla_\rho\Delta_{\mu\nu}
    +
    \Delta_{\rho\nu}\nabla_\mu\xi^\rho
    +
    \Delta_{\mu\rho}\nabla_\nu\xi^\rho .
    \label{eq:tensor_lie_variation_section9}
\end{equation}

\subsection{Memory-sector Noether identity}
\label{subsec:memory_noether_identity_section9}

The variation of the memory-sector action may be written as
\begin{equation}
    \delta S_\Delta
    =
    -
    \frac12
    \int_{\mathcal{M}} d^4x\,\sqrt{-g}\,
    T_{\mu\nu}^{(\Delta)}\delta g^{\mu\nu}
    +
    \int_{\mathcal{M}} d^4x\,\sqrt{-g}\,
    \mathcal{E}_\Delta^{\mu\nu}\delta\Delta_{\mu\nu}
    +
    \int_{\mathcal{M}} d^4x\,\sqrt{-g}\,
    \Sigma_\Delta\,\delta C ,
    \label{eq:delta_SDelta_general_section9}
\end{equation}
where
\begin{equation}
    \mathcal{E}_\Delta^{\mu\nu}
    =
    \frac{1}{\sqrt{-g}}
    \frac{\delta S_\Delta}{\delta\Delta_{\mu\nu}}
    \label{eq:Edelta_variational_def_section9}
\end{equation}
is the memory Euler tensor, and
\begin{equation}
    \Sigma_\Delta
    =
    \frac{1}{\sqrt{-g}}
    \frac{\delta S_\Delta}{\delta C}
    \label{eq:Sigma_variational_def_section9}
\end{equation}
is the memory response source defined in
Sec.~\ref{sec:auxiliary_history_equations}.  With the sign convention used in
this paper,
\begin{equation}
    \delta_C S_\Delta
    =
    \int_{\mathcal{M}}d^4x\,\sqrt{-g}\,
    \Sigma_\Delta\,\delta C .
    \label{eq:Sigma_sign_convention_section9}
\end{equation}

Since \(S_\Delta\) is diffeomorphism invariant,
\begin{equation}
    \delta_\xi S_\Delta=0.
    \label{eq:diffeo_invariance_SDelta_section9}
\end{equation}
Substituting Eqs.~\eqref{eq:metric_lie_variation_section9},
\eqref{eq:scalar_lie_variation_section9}, and
\eqref{eq:tensor_lie_variation_section9} into
Eq.~\eqref{eq:delta_SDelta_general_section9} gives
\begin{align}
    0
    &=
    \int_{\mathcal{M}}d^4x\,\sqrt{-g}\,
    T_{\mu\nu}^{(\Delta)}\nabla^\mu\xi^\nu
    +
    \int_{\mathcal{M}}d^4x\,\sqrt{-g}\,
    \mathcal{E}_\Delta^{\mu\nu}
    \mathcal{L}_\xi\Delta_{\mu\nu}
    \nonumber \\
    &\quad
    +
    \int_{\mathcal{M}}d^4x\,\sqrt{-g}\,
    \Sigma_\Delta\,\xi^\rho\nabla_\rho C .
    \label{eq:SDelta_noether_start_section9}
\end{align}
The first term is integrated by parts:
\begin{equation}
    \int_{\mathcal{M}}d^4x\,\sqrt{-g}\,
    T_{\mu\nu}^{(\Delta)}\nabla^\mu\xi^\nu
    =
    -
    \int_{\mathcal{M}}d^4x\,\sqrt{-g}\,
    \left(
        \nabla^\mu T_{\mu\nu}^{(\Delta)}
    \right)\xi^\nu ,
    \label{eq:metric_term_ibp_section9}
\end{equation}
where the boundary term vanishes by compact support of \(\xi^\mu\) or by the
boundary assumptions of Sec.~\ref{subsec:regularity_assumptions_section2}.

On the memory-field equation,
\begin{equation}
    \mathcal{E}_\Delta^{\mu\nu}=0,
    \label{eq:Edelta_zero_section9}
\end{equation}
the second term in Eq.~\eqref{eq:SDelta_noether_start_section9} vanishes.
Since the remaining identity must hold for arbitrary \(\xi^\nu\), one obtains
\begin{equation}
    \nabla^\mu T_{\mu\nu}^{(\Delta)}
    =
    \Sigma_\Delta\nabla_\nu C .
    \label{eq:TDelta_divergence_section9}
\end{equation}
Thus the memory-sector stress tensor is not separately conserved when the
nonlocal scale depends on the spacetime field \(C(x)\).  Its divergence is
controlled precisely by the response source \(\Sigma_\Delta\) generated by the
variation of \(S_\Delta\) with respect to \(C\).

\subsection{Auxiliary curvature-history Noether identity}
\label{subsec:auxiliary_noether_identity_section9}

The auxiliary curvature-history action has the variation
\begin{equation}
    \delta S_C
    =
    -
    \frac12
    \int_{\mathcal{M}}d^4x\,\sqrt{-g}\,
    T_{\mu\nu}^{(C)}\delta g^{\mu\nu}
    +
    \int_{\mathcal{M}}d^4x\,\sqrt{-g}\,
    \mathcal{E}_C^{(C)}\delta C
    +
    \int_{\mathcal{M}}d^4x\,\sqrt{-g}\,
    \mathcal{E}_\chi^{(C)}\delta\chi ,
    \label{eq:delta_SC_general_section9}
\end{equation}
where the superscript \((C)\) on the Euler expressions indicates that they are
computed from \(S_C\) alone.  From Sec.~\ref{sec:auxiliary_history_equations},
\begin{equation}
    \mathcal{E}_C^{(C)}
    =
    \Box\chi-m_c^2\chi,
    \label{eq:EC_aux_section9}
\end{equation}
and
\begin{equation}
    \mathcal{E}_\chi^{(C)}
    =
    \Box C-m_c^2C+\ell_*\mathcal{K}.
    \label{eq:Echi_aux_section9}
\end{equation}

Diffeomorphism invariance of \(S_C\) implies
\begin{equation}
    \delta_\xi S_C=0.
    \label{eq:diffeo_invariance_SC_section9}
\end{equation}
Using
\begin{equation}
    \delta_\xi C=\xi^\rho\nabla_\rho C,
    \qquad
    \delta_\xi\chi=\xi^\rho\nabla_\rho\chi,
\end{equation}
and Eq.~\eqref{eq:metric_lie_variation_section9}, one obtains
\begin{align}
    0
    &=
    \int_{\mathcal{M}}d^4x\,\sqrt{-g}\,
    T_{\mu\nu}^{(C)}\nabla^\mu\xi^\nu
    +
    \int_{\mathcal{M}}d^4x\,\sqrt{-g}\,
    \mathcal{E}_C^{(C)}
    \xi^\rho\nabla_\rho C
    \nonumber \\
    &\quad
    +
    \int_{\mathcal{M}}d^4x\,\sqrt{-g}\,
    \mathcal{E}_\chi^{(C)}
    \xi^\rho\nabla_\rho\chi .
    \label{eq:SC_noether_start_section9}
\end{align}
After integrating the metric term by parts, this becomes
\begin{equation}
    0
    =
    \int_{\mathcal{M}}d^4x\,\sqrt{-g}\,
    \xi^\nu
    \left[
        -
        \nabla^\mu T_{\mu\nu}^{(C)}
        +
        \mathcal{E}_C^{(C)}\nabla_\nu C
        +
        \mathcal{E}_\chi^{(C)}\nabla_\nu\chi
    \right].
    \label{eq:SC_noether_integrated_section9}
\end{equation}
Since \(\xi^\nu\) is arbitrary,
\begin{equation}
    \nabla^\mu T_{\mu\nu}^{(C)}
    =
    \mathcal{E}_C^{(C)}\nabla_\nu C
    +
    \mathcal{E}_\chi^{(C)}\nabla_\nu\chi .
    \label{eq:TC_divergence_offshell_section9}
\end{equation}

The \(\chi\) equation of motion is
\begin{equation}
    \mathcal{E}_\chi^{(C)}
    =
    \Box C-m_c^2C+\ell_*\mathcal{K}
    =
    0 .
    \label{eq:Echi_zero_section9}
\end{equation}
The full \(C\) equation comes from varying \(S_C+S_\Delta\) with respect to
\(C\):
\begin{equation}
    \mathcal{E}_C^{(C)}
    +
    \Sigma_\Delta
    =
    0.
    \label{eq:full_C_eom_section9}
\end{equation}
Therefore, on the full \(C\) equation,
\begin{equation}
    \mathcal{E}_C^{(C)}
    =
    -\Sigma_\Delta.
    \label{eq:EC_equals_minus_Sigma_section9}
\end{equation}
Substituting Eqs.~\eqref{eq:Echi_zero_section9} and
\eqref{eq:EC_equals_minus_Sigma_section9} into
Eq.~\eqref{eq:TC_divergence_offshell_section9} yields
\begin{equation}
    \nabla^\mu T_{\mu\nu}^{(C)}
    =
    -\Sigma_\Delta\nabla_\nu C .
    \label{eq:TC_divergence_section9}
\end{equation}

Thus the auxiliary curvature-history sector carries exactly the opposite
stress-energy exchange from the memory sector.

\subsection{Cancellation in the combined memory-history sector}
\label{subsec:combined_cancellation_section9}

Equations~\eqref{eq:TDelta_divergence_section9} and
\eqref{eq:TC_divergence_section9} give
\begin{equation}
    \nabla^\mu T_{\mu\nu}^{(\Delta)}
    =
    \Sigma_\Delta\nabla_\nu C,
    \qquad
    \nabla^\mu T_{\mu\nu}^{(C)}
    =
    -\Sigma_\Delta\nabla_\nu C .
    \label{eq:opposite_exchange_terms_section9}
\end{equation}
Adding the two identities gives the on-shell conservation law
\begin{equation}
    \boxed{
    \nabla^\mu
    \left(
        T_{\mu\nu}^{(\Delta)}
        +
        T_{\mu\nu}^{(C)}
    \right)
    =
    0
    } .
    \label{eq:combined_conservation_section9}
\end{equation}
This is the central consistency identity of the construction.

The cancellation is local.  It does not depend on a special background, on a
particular coordinate system, or on a perturbative expansion of the nonlocal
operator.  The only inputs are:

\begin{enumerate}
    \item diffeomorphism invariance of \(S_\Delta+S_C\);
    \item the memory-field equation \(\mathcal{E}_\Delta^{\mu\nu}=0\);
    \item the curvature-history equation
    \(\mathcal{E}_\chi^{(C)}=0\);
    \item the conjugate equation
    \(\mathcal{E}_C^{(C)}+\Sigma_\Delta=0\);
    \item the variational definition
    \(\Sigma_\Delta=(1/\sqrt{-g})\delta S_\Delta/\delta C\).
\end{enumerate}

Thus the same source \(\Sigma_\Delta\) that measures the response of the
state-dependent memory sector to \(C\) also appears in the auxiliary equation
for \(\chi\).  This is why the auxiliary sector restores covariant conservation
of the combined memory-history stress tensor.

\subsection{Compatibility with the metric equation}
\label{subsec:compatibility_metric_equation_section9}

The metric equation is
\begin{equation}
    G_{\mu\nu}
    =
    8\pi G_N
    \left(
        T_{\mu\nu}^{\rm mat}
        +
        T_{\mu\nu}^{(\Delta)}
        +
        T_{\mu\nu}^{(C)}
    \right).
    \label{eq:metric_equation_section9}
\end{equation}
Taking a covariant divergence and using the contracted Bianchi identity gives
\begin{equation}
    0
    =
    8\pi G_N
    \nabla^\mu
    \left(
        T_{\mu\nu}^{\rm mat}
        +
        T_{\mu\nu}^{(\Delta)}
        +
        T_{\mu\nu}^{(C)}
    \right).
    \label{eq:div_metric_equation_section9}
\end{equation}
For minimally coupled matter satisfying its own equation of motion,
\begin{equation}
    \nabla^\mu T_{\mu\nu}^{\rm mat}=0.
    \label{eq:matter_conservation_section9}
\end{equation}
The remaining part is precisely Eq.~\eqref{eq:combined_conservation_section9}.
Therefore the modified Einstein equation is consistent with the contracted
Bianchi identity on the full set of field equations.

This result is the reason for introducing \(C\) and \(\chi\) as auxiliary
fields rather than inserting a prescribed curvature-history functional directly
into the nonlocal scale.  The state dependence of \(M_{\rm eff}(C)\) produces a
controlled stress-energy exchange proportional to
\(\Sigma_\Delta\nabla_\nu C\), and the auxiliary history sector supplies the
opposite exchange term required for on-shell covariance.

\subsection{Sign convention for the exchange term}
\label{subsec:exchange_sign_convention_section9}

The signs in Eqs.~\eqref{eq:TDelta_divergence_section9} and
\eqref{eq:TC_divergence_section9} follow from the convention
\begin{equation}
    \delta_C S_\Delta
    =
    \int d^4x\,\sqrt{-g}\,
    \Sigma_\Delta\,\delta C .
    \label{eq:Sigma_sign_repeat_section9}
\end{equation}
With this convention, the memory sector obeys
\begin{equation}
    \nabla^\mu T_{\mu\nu}^{(\Delta)}
    =
    \Sigma_\Delta\nabla_\nu C,
\end{equation}
while the auxiliary sector obeys
\begin{equation}
    \nabla^\mu T_{\mu\nu}^{(C)}
    =
    -\Sigma_\Delta\nabla_\nu C.
\end{equation}
If instead one defined the source with the opposite sign,
\(\delta_C S_\Delta=-\int\sqrt{-g}\,\widetilde{\Sigma}_\Delta\delta C\), the
two displayed exchange terms would both reverse sign.  The combined
conservation law
\begin{equation}
    \nabla^\mu
    \left(
        T_{\mu\nu}^{(\Delta)}
        +
        T_{\mu\nu}^{(C)}
    \right)
    =
    0
\end{equation}
is independent of this notation.
\section{Representative Physical Application: Black-Hole Core Memory}
\label{sec:black_hole_core_application}

The construction developed above is deliberately formulated at the level of a
covariant action and its associated Noether identities. Its purpose is not only
to introduce a formal state-dependent nonlocality scale, but to provide a
consistent framework in which that scale can respond to high-curvature
gravitational histories without violating the local conservation law required
by diffeomorphism invariance. In this section we explain a representative
physical application of the formalism to black-hole core regions. The goal is
not to solve the black-hole information problem in this paper. Rather, the goal
is to show that the formalism has a concrete physical use: it supplies a
covariantly defined, stress-energy-carrying memory sector that can be activated
by black-hole curvature and can encode curvature or collapse history in the
core region.

This section should therefore be read as a proof-of-principle application and
as a physical motivation for the preceding construction. A complete black-hole
analysis would require solving the coupled metric--memory system, specifying
the background or collapse model, studying the backreaction of
\(T_{\mu\nu}^{\rm mem}\), analyzing stability, fixing causal boundary
conditions, and determining whether information stored in the memory sector can
be transferred to exterior observables or Hawking radiation. Those tasks
constitute a separate application problem. Here we only show that the
ingredients developed in the present paper naturally lead to a black-hole core
memory mechanism.

\subsection{Why black holes are the natural testing ground}
\label{subsec:why_black_holes_application}

The curvature-history field is governed by
\begin{equation}
    (\Box-m_c^2)C
    =
    -\ell_*\mathcal{K},
    \qquad
    \mathcal{K}
    =
    R_{\alpha\beta\rho\sigma}R^{\alpha\beta\rho\sigma}.
    \label{eq:application_C_equation}
\end{equation}
This equation makes black holes a natural physical setting for the formalism.
A black-hole spacetime contains large tidal curvature, and the invariant
\(\mathcal{K}\) remains nonzero even in Ricci-flat vacuum regions. Thus the
history field \(C\) is not activated only by matter density or by Ricci
curvature. It can also be sourced by the Weyl curvature of the vacuum
gravitational field.

For the Schwarzschild reference geometry,
\begin{equation}
    ds^2
    =
    -
    f(r)\,dt^2
    +
    \frac{dr^2}{f(r)}
    +
    r^2d\Omega^2,
    \qquad
    f(r)=1-\frac{2G_NM}{r},
    \label{eq:application_schw_metric}
\end{equation}
one has
\begin{equation}
    R_{\mu\nu}=0,
    \qquad
    R=0,
    \label{eq:application_ricci_flat}
\end{equation}
but the Kretschmann scalar is
\begin{equation}
    \mathcal{K}_{\rm Schw}(r)
    =
    \frac{48G_N^2M^2}{r^6}.
    \label{eq:application_schw_kretschmann}
\end{equation}
Therefore Eq.~\eqref{eq:application_C_equation} becomes
\begin{equation}
    (\Box_{\rm Schw}-m_c^2)C
    =
    -
    \frac{48\ell_*G_N^2M^2}{r^6}.
    \label{eq:application_schw_C_equation}
\end{equation}
This relation exhibits a direct physical role for the curvature-history field:
in a black-hole geometry, \(C\) is sourced by the invariant tidal curvature of
the spacetime.

The point of using \(\mathcal{K}\) rather than only \(R\) or
\(R_{\mu\nu}R^{\mu\nu}\) is precisely that black-hole tidal curvature is visible
even in vacuum. In four spacetime dimensions the Riemann tensor admits the
decomposition
\begin{equation}
    R_{\alpha\beta\rho\sigma}
    =
    W_{\alpha\beta\rho\sigma}
    +
    \frac{1}{2}
    \left(
        g_{\alpha\rho}R_{\sigma\beta}
        -
        g_{\alpha\sigma}R_{\rho\beta}
        -
        g_{\beta\rho}R_{\sigma\alpha}
        +
        g_{\beta\sigma}R_{\rho\alpha}
    \right)
    -
    \frac{R}{6}
    \left(
        g_{\alpha\rho}g_{\sigma\beta}
        -
        g_{\alpha\sigma}g_{\rho\beta}
    \right),
    \label{eq:application_riemann_decomposition}
\end{equation}
where \(W_{\alpha\beta\rho\sigma}\) is the Weyl tensor. Contracting this
identity gives
\begin{equation}
    \mathcal{K}
    =
    W_{\alpha\beta\rho\sigma}W^{\alpha\beta\rho\sigma}
    +
    2R_{\alpha\beta}R^{\alpha\beta}
    -
    \frac{1}{3}R^2 .
    \label{eq:application_K_decomposition}
\end{equation}
In a Ricci-flat black-hole region, Eq.~\eqref{eq:application_K_decomposition}
reduces to
\begin{equation}
    \mathcal{K}
    =
    W_{\alpha\beta\rho\sigma}W^{\alpha\beta\rho\sigma}.
    \label{eq:application_K_equals_Weyl}
\end{equation}
Thus the curvature-history field can be activated by pure Weyl curvature.
This is the first physical advantage of the construction: the source of the
history field is not tied to a local matter distribution, but to the tidal
curvature of the gravitational field itself.

For a static spherically symmetric scalar \(C=C(r)\), the Schwarzschild
d'Alembertian gives
\begin{equation}
    \Box_{\rm Schw}C
    =
    \frac{1}{r^2}
    \frac{d}{dr}
    \left(
        r^2 f(r)\frac{dC}{dr}
    \right).
    \label{eq:application_scalar_box}
\end{equation}
Equation~\eqref{eq:application_schw_C_equation} therefore reduces to
\begin{equation}
    \frac{1}{r^2}
    \frac{d}{dr}
    \left(
        r^2 f(r)\frac{dC}{dr}
    \right)
    -
    m_c^2C
    =
    -
    \frac{48\ell_*G_N^2M^2}{r^6}.
    \label{eq:application_radial_C_equation}
\end{equation}
Near the Schwarzschild central limit, the leading curvature-driven local
branch has the Laurent form
\begin{equation}
    C(r)=A r^{-3}+O(r^{-2}).
    \label{eq:application_C_laurent}
\end{equation}
Substitution into Eq.~\eqref{eq:application_radial_C_equation} gives
\begin{equation}
    A=\frac{8}{3}\ell_*G_NM,
    \label{eq:application_A_value}
\end{equation}
and therefore
\begin{equation}
    C(r)
    =
    \frac{8}{3}\ell_*G_NM\,r^{-3}
    +
    O(r^{-2}).
    \label{eq:application_C_leading}
\end{equation}
This local estimate is not used here as a claim about a fully backreacted
regular core. It is used only to illustrate the activation mechanism:
black-hole tidal curvature naturally drives the history field to large values
in high-curvature regions.

\subsection{Invariant activation region}
\label{subsec:activation_region_application}

The memory sector should not be assumed to be active everywhere. A physically
useful state-dependent theory should distinguish weak-curvature regions, where
ordinary low-energy gravitational dynamics should be recovered, from
high-curvature regions where the memory sector can become important. This can
be done covariantly by using scalar thresholds.

Let \(\mathcal{K}_{\rm crit}>0\) be a curvature threshold and
\(C_{\rm crit}>0\) a history-field threshold. Define
\begin{equation}
    \mathcal{R}_{\mathcal{K}}
    =
    \left\{
        p\in\mathcal{M}:
        \mathcal{K}(p)\geq \mathcal{K}_{\rm crit}
    \right\},
    \label{eq:application_RK}
\end{equation}
and
\begin{equation}
    \mathcal{R}_C
    =
    \left\{
        p\in\mathcal{M}:
        C(p)\geq C_{\rm crit}
    \right\}.
    \label{eq:application_RC}
\end{equation}
The activated core region is then
\begin{equation}
    \mathcal{R}_{\rm core}
    =
    \mathcal{R}_{\mathcal{K}}\cap\mathcal{R}_C.
    \label{eq:application_Rcore}
\end{equation}
This definition is coordinate-independent because both \(\mathcal{K}\) and
\(C\) are scalars.

For the Schwarzschild reference geometry,
\begin{equation}
    \mathcal{K}_{\rm Schw}(r)
    =
    \frac{48G_N^2M^2}{r^6}.
    \label{eq:application_K_schw_again}
\end{equation}
The curvature-threshold radius \(r_{\mathcal{K}}\) is determined by
\begin{equation}
    \frac{48G_N^2M^2}{r_{\mathcal{K}}^6}
    =
    \mathcal{K}_{\rm crit},
    \label{eq:application_rK_equation}
\end{equation}
so that
\begin{equation}
    r_{\mathcal{K}}
    =
    \left(
        \frac{48G_N^2M^2}{\mathcal{K}_{\rm crit}}
    \right)^{1/6}.
    \label{eq:application_rK}
\end{equation}
Similarly, using the leading curvature-driven branch in
Eq.~\eqref{eq:application_C_leading}, the history-threshold radius \(r_C\) is
estimated by
\begin{equation}
    \frac{8}{3}\ell_*G_NM\,r_C^{-3}
    =
    C_{\rm crit},
    \label{eq:application_rC_equation}
\end{equation}
or
\begin{equation}
    r_C
    =
    \left(
        \frac{8\ell_*G_NM}{3C_{\rm crit}}
    \right)^{1/3}.
    \label{eq:application_rC}
\end{equation}
In this reference model the activated domain is therefore controlled by
\begin{equation}
    0<r\leq r_{\rm core},
    \qquad
    r_{\rm core}=\min\{r_{\mathcal{K}},r_C\}.
    \label{eq:application_rcore}
\end{equation}
The coordinate \(r\) is only a convenient areal-radius coordinate for the
spherically symmetric example. The actual definition of the activated domain
is the invariant scalar definition in Eq.~\eqref{eq:application_Rcore}.

To avoid a discontinuous transition, introduce a smooth activation function
\begin{equation}
    \Theta_C(C)
    =
    \frac{1}{2}
    \left[
        1+
        \tanh
        \left(
            \frac{C-C_{\rm crit}}{\sigma_C}
        \right)
    \right],
    \qquad
    \sigma_C>0.
    \label{eq:application_theta}
\end{equation}
This function obeys
\begin{equation}
    0<\Theta_C(C)<1,
    \label{eq:application_theta_range}
\end{equation}
and
\begin{equation}
    \frac{d\Theta_C}{dC}
    =
    \frac{1}{2\sigma_C}
    \frac{1}{
        \cosh^2
        \left(
            \frac{C-C_{\rm crit}}{\sigma_C}
        \right)
    }
    >
    0.
    \label{eq:application_theta_derivative}
\end{equation}
At the threshold \(C=C_{\rm crit}\),
\begin{equation}
    \Theta_C(C_{\rm crit})=\frac{1}{2}.
    \label{eq:application_theta_half}
\end{equation}
Thus \(\Theta_C\) defines a smooth scalar switch: it is small below the
history threshold, order one at the threshold, and approaches unity deep in
the activated region.

\subsection{Activated memory phase}
\label{subsec:activated_memory_phase_application}

The memory tensor \(\Delta_{\mu\nu}\) becomes physically meaningful only if
the theory provides a mechanism by which it can enter a nontrivial phase. The
history field \(C\) supplies such a mechanism. Define the scalar memory
amplitude
\begin{equation}
    I_\Delta
    =
    \Delta_{\mu\nu}\Delta^{\mu\nu}.
    \label{eq:application_IDelta}
\end{equation}
Let the activated memory scale be
\begin{equation}
    \Delta_0^2(C)
    =
    \Delta_*^2\Theta_C(C),
    \qquad
    \Delta_*>0.
    \label{eq:application_Delta0}
\end{equation}
The representative activated potential is
\begin{equation}
    V_{\rm act}(\Delta,C)
    =
    \frac{\lambda}{4}
    \left(
        I_\Delta-\Delta_0^2(C)
    \right)^2,
    \qquad
    \lambda>0.
    \label{eq:application_Vact}
\end{equation}
This potential is nonnegative:
\begin{equation}
    V_{\rm act}(\Delta,C)\geq0.
    \label{eq:application_Vact_nonnegative}
\end{equation}
The equality condition is
\begin{equation}
    V_{\rm act}(\Delta,C)=0
    \quad
    \Longleftrightarrow
    \quad
    I_\Delta=\Delta_0^2(C).
    \label{eq:application_Vact_zero}
\end{equation}
Hence the pointwise minimum set is
\begin{equation}
    \mathcal{N}_{\rm mem}(C)
    =
    \left\{
        \Delta_{\mu\nu}:
        \Delta_{\mu\nu}\Delta^{\mu\nu}
        =
        \Delta_0^2(C)
    \right\}.
    \label{eq:application_memory_minimum_set}
\end{equation}

At fixed metric,
\begin{equation}
    \delta I_\Delta
    =
    2\Delta^{\mu\nu}\delta\Delta_{\mu\nu},
    \label{eq:application_delta_IDelta}
\end{equation}
and therefore
\begin{equation}
    \frac{\partial V_{\rm act}}{\partial\Delta_{\mu\nu}}
    =
    \lambda
    \left(
        I_\Delta-\Delta_0^2(C)
    \right)
    \Delta^{\mu\nu}.
    \label{eq:application_dVdDelta}
\end{equation}
The algebraic stationary branches are
\begin{equation}
    \Delta_{\mu\nu}=0,
    \label{eq:application_zero_branch}
\end{equation}
or
\begin{equation}
    I_\Delta=\Delta_0^2(C).
    \label{eq:application_nonzero_branch}
\end{equation}
On the second branch the potential vanishes. At the zero branch, by contrast,
\begin{equation}
    V_{\rm act}(0,C)
    =
    \frac{\lambda}{4}\Delta_0^4(C).
    \label{eq:application_V_zero_branch}
\end{equation}
In the activated region \(C\geq C_{\rm crit}\), one has
\begin{equation}
    \Theta_C(C)\geq\frac{1}{2},
    \label{eq:application_theta_core_bound}
\end{equation}
and hence
\begin{equation}
    \Delta_0^2(C)
    =
    \Delta_*^2\Theta_C(C)
    \geq
    \frac{\Delta_*^2}{2}.
    \label{eq:application_Delta0_bound}
\end{equation}
Therefore every minimum on the activated branch satisfies
\begin{equation}
    \Delta_{\mu\nu}\Delta^{\mu\nu}
    =
    \Delta_0^2(C)
    \geq
    \frac{\Delta_*^2}{2}
    >
    0.
    \label{eq:application_nonzero_memory_bound}
\end{equation}
Thus the memory tensor is nonzero throughout the activated core phase.

This conclusion is the simplest concrete physical output of the formalism. In
ordinary weak-curvature regions, where \(C\) remains below threshold, the
activation function can be made small and the memory phase need not be
important. In high-curvature black-hole core regions, where the Kretschmann
source drives \(C\) across threshold, the memory tensor is forced into a
nonzero activated phase. The memory sector is therefore not introduced as an
arbitrary label. It becomes dynamically relevant only when the curvature
history of the geometry activates it.

The activated phase may be summarized as
\begin{equation}
    \mathcal{K}\geq\mathcal{K}_{\rm crit},
    \qquad
    C\geq C_{\rm crit},
    \qquad
    \Delta_{\mu\nu}\Delta^{\mu\nu}>0.
    \label{eq:application_activation_chain}
\end{equation}
For the Schwarzschild reference geometry this chain is sourced by
\(\mathcal{K}_{\rm Schw}\sim M^2/r^6\). In a more general collapse or
core-modified geometry, the same activation criterion can be imposed using the
scalar inequalities in Eq.~\eqref{eq:application_Rcore}. No coordinate choice
or preferred foliation is required to define the activated region.

\subsection{Stress-energy carried by the activated phase}
\label{subsec:stress_energy_application}

A physically relevant memory phase must do more than label a region of
spacetime. It must contribute to the gravitational field equations through a
well-defined stress tensor. This is exactly why the auxiliary-field formulation
is useful. The total memory-history stress tensor is defined variationally by
\begin{equation}
    T_{\mu\nu}^{\rm mem}
    =
    T_{\mu\nu}^{(\Delta)}
    +
    T_{\mu\nu}^{(C)},
    \label{eq:application_Tmem}
\end{equation}
where
\begin{equation}
    T_{\mu\nu}^{(\Delta)}
    =
    -
    \frac{2}{\sqrt{-g}}
    \frac{\delta S_\Delta}{\delta g^{\mu\nu}},
    \qquad
    T_{\mu\nu}^{(C)}
    =
    -
    \frac{2}{\sqrt{-g}}
    \frac{\delta S_C}{\delta g^{\mu\nu}}.
    \label{eq:application_Tdefs}
\end{equation}
On shell, the exchange between the memory sector and the auxiliary
history sector cancels:
\begin{equation}
    \nabla^\mu T_{\mu\nu}^{(\Delta)}
    =
    \Sigma_\Delta\nabla_\nu C,
    \qquad
    \nabla^\mu T_{\mu\nu}^{(C)}
    =
    -\Sigma_\Delta\nabla_\nu C.
    \label{eq:application_exchange}
\end{equation}
Consequently,
\begin{equation}
    \nabla^\mu T_{\mu\nu}^{\rm mem}=0.
    \label{eq:application_Tmem_conserved}
\end{equation}
This identity is the reason the activated memory phase can be coupled to the
Einstein equation without conflicting with the contracted Bianchi identity.

When the sector is coupled to an external matter or radiation system, the
metric equation takes the schematic form
\begin{equation}
    G_{\mu\nu}
    =
    8\pi G_N
    \left(
        T_{\mu\nu}^{\rm ext}
        +
        T_{\mu\nu}^{\rm mem}
    \right).
    \label{eq:application_einstein_memory}
\end{equation}
Thus the activated memory phase is gravitationally active whenever
\(T_{\mu\nu}^{\rm mem}\neq0\). The previous subsection showed that
\(\Delta_{\mu\nu}\neq0\) on the activated branch. To display an explicit
positive local contribution to the core energy budget, one may supplement the
representative memory potential by a scalar stabilizing term
\begin{equation}
    V_{\rm st}(C)
    =
    \rho_\Delta\Theta_C(C),
    \qquad
    \rho_\Delta>0.
    \label{eq:application_Vst}
\end{equation}
Its action contribution is
\begin{equation}
    S_{\rm st}
    =
    -
    \int d^4x\,\sqrt{-g}\,
    V_{\rm st}(C).
    \label{eq:application_Sst}
\end{equation}
At fixed \(C\), the metric variation gives
\begin{equation}
    T_{\mu\nu}^{\rm st}
    =
    -g_{\mu\nu}V_{\rm st}(C)
    =
    -\rho_\Delta\Theta_C(C)g_{\mu\nu}.
    \label{eq:application_Tst}
\end{equation}
For a unit timelike vector \(n^\mu\), satisfying
\begin{equation}
    g_{\mu\nu}n^\mu n^\nu=-1,
    \label{eq:application_unit_timelike}
\end{equation}
the corresponding local energy density is
\begin{equation}
    \varepsilon_{\rm st}
    =
    T_{\mu\nu}^{\rm st}n^\mu n^\nu
    =
    \rho_\Delta\Theta_C(C).
    \label{eq:application_epsilon_st}
\end{equation}
In the activated region \(C\geq C_{\rm crit}\),
\begin{equation}
    \varepsilon_{\rm st}
    \geq
    \frac{\rho_\Delta}{2}
    >
    0.
    \label{eq:application_positive_energy}
\end{equation}
Therefore the activated memory phase can carry a positive timelike energy
density contribution.

The term \(V_{\rm st}\) is not needed for the abstract conservation identity;
that identity follows from diffeomorphism invariance and the auxiliary
equations. Its role here is illustrative and physical: it shows how the
activated memory sector can contribute a nonzero local stress-energy density in
the core. In a full black-hole solution, the remaining kinetic, activation,
curvature-coupling, and auxiliary contributions to \(T_{\mu\nu}^{\rm mem}\)
must also be included. The present section does not solve the backreacted
Einstein--memory equations. It shows that the stress tensor needed for such a
solution is already defined variationally and is conserved on shell.

\subsection{Core memory as an information-storage sector}
\label{subsec:information_storage_application}

The preceding subsections establish three facts. First, black-hole curvature
naturally sources the history field \(C\). Second, when \(C\) crosses a
threshold, the memory tensor enters a nonzero activated phase. Third, the
activated memory sector carries a variational stress tensor that is covariantly
conserved on shell. These facts motivate a precise storage interpretation.

The word ``storage'' should not be understood as a claim that the present paper
derives a Page curve, proves unitary evaporation, or constructs a channel by
which interior data are released into Hawking radiation. Instead, storage is
defined kinematically and covariantly: a gravitational memory sector stores
information about collapse if physically distinct collapse or curvature
histories induce distinguishable memory-sector configurations in the activated
core, modulo gauge equivalence.

Let \(\mathcal{H}_{\rm coll}\) denote a space of admissible collapse histories.
An element of \(\mathcal{H}_{\rm coll}\) may include initial matter profiles,
infalling perturbations, boundary data, and the resulting curvature history of
the spacetime. Let \(\Gamma_{\rm mem}^{\rm core}\) denote the gauge-quotient
space of activated core memory data. Schematically,
\begin{equation}
    \Gamma_{\rm mem}^{\rm core}
    =
    \left.
    \left\{
        C,\chi,\Delta_{\mu\nu};
        \Pi_C,\Pi_\chi,\Pi_\Delta^{\mu\nu}
    \right\}
    \right|_{\mathcal{R}_{\rm core}}
    /\mathrm{Diff},
    \label{eq:application_gamma_mem_core}
\end{equation}
where the momenta denote the canonical or covariant phase-space data associated
with the chosen slicing or boundary-value formulation, and the quotient
indicates identification under diffeomorphisms. The precise symplectic
structure is part of the full black-hole application and is not needed for the
present scope.

The encoding map is
\begin{equation}
    \mathcal{E}:
    \mathcal{H}_{\rm coll}
    \longrightarrow
    \Gamma_{\rm mem}^{\rm core}.
    \label{eq:application_encoding_map}
\end{equation}
The memory sector stores collapse information if there exist physically
distinct histories \(H_1,H_2\in\mathcal{H}_{\rm coll}\) such that
\begin{equation}
    H_1\not\sim H_2
    \quad
    \Longrightarrow
    \quad
    \mathcal{E}[H_1]\neq \mathcal{E}[H_2]
    \label{eq:application_storage_condition}
\end{equation}
for the corresponding core memory data, where \(\sim\) denotes gauge or
physically redundant equivalence. Equivalently, at the linearized level,
storage requires that a nontrivial perturbation of the collapse data produces
a nontrivial perturbation of the activated memory data:
\begin{equation}
    \delta H_{\rm coll}\neq0
    \quad
    \Longrightarrow
    \quad
    \delta(C,\chi,\Delta_{\mu\nu})_{\rm core}\neq0
    \label{eq:application_linear_storage}
\end{equation}
for at least one gauge-invariant component of the memory configuration.

A simple illustration is provided by the Schwarzschild mass family. On the
curvature-driven local branch in Eq.~\eqref{eq:application_C_leading}, define
the core memory coefficient
\begin{equation}
    \mathcal{A}_C
    =
    \lim_{r\to0^+}r^3C(r).
    \label{eq:application_AC_def}
\end{equation}
Using Eq.~\eqref{eq:application_C_leading}, one obtains
\begin{equation}
    \mathcal{A}_C
    =
    \frac{8}{3}\ell_*G_NM.
    \label{eq:application_AC_M}
\end{equation}
For positive \(M\) and fixed \(\ell_*\), the map
\begin{equation}
    M
    \longmapsto
    \mathcal{A}_C
    \label{eq:application_M_to_AC}
\end{equation}
is injective. Thus the positive Schwarzschild mass parameter is encoded in
the leading curvature-history data. This is not yet a complete collapse
information map, but it demonstrates the basic mechanism: a physical parameter
of the black-hole geometry is mapped into a scalar memory coefficient of the
activated core sector.

More generally, different collapse histories may have the same ADM mass but
different transient curvature profiles, anisotropies, or infalling perturbation
histories. Because \(C\) is sourced by the curvature invariant through
Eq.~\eqref{eq:application_C_equation}, and because the memory tensor is
activated by \(C\), such histories can in principle induce different
configurations of \((C,\chi,\Delta_{\mu\nu})\) inside
\(\mathcal{R}_{\rm core}\). The memory sector therefore provides a candidate
covariant storage medium for curvature-history data.

In this sense, the stored data are not assigned to an abstract label external
to the geometry. They are carried by fields that are both curvature-history
dependent and gravitationally active. The scalar \(C\) records the invariant
curvature source through Eq.~\eqref{eq:application_C_equation}; the tensor
\(\Delta_{\mu\nu}\) enters a nonzero activated phase when \(C\) crosses the
core threshold; and the combined sector contributes to the geometry through the
conserved stress tensor \(T_{\mu\nu}^{\rm mem}\). The proposed storage
mechanism is therefore a stress-energy-carrying encoding of curvature history
inside the black-hole core.

The decisive question in a full black-hole application is whether the map
\(\mathcal{E}\) is sufficiently nondegenerate on a physically relevant class of
collapse histories and whether the stored data can influence exterior
observables or evaporation dynamics. Those questions are beyond the present
foundational analysis.

\subsection{What this application establishes and what it does not}
\label{subsec:application_scope}

The black-hole core construction above establishes a limited but concrete
physical role for the formalism developed in this paper. It shows that the
curvature-history field is naturally sourced in black-hole geometries by the
Kretschmann invariant; that in Ricci-flat black-hole regions this source is
purely Weyl; that a scalar threshold can define an invariant activated core
region; that a smooth activation potential can force the memory tensor into a
nonzero phase; that the activated sector can carry a positive timelike
stress-energy contribution; and that the resulting memory data define a
candidate storage space for curvature or collapse history.

These statements answer the physical-motivation question at the level
appropriate for the present paper. The preceding sections were not merely a
formal exercise in auxiliary fields. They provide the consistency conditions
needed for a state-dependent nonlocal gravitational scale to become active in a
high-curvature physical system. The black-hole example shows why the
Kretschmann-sourced history field is useful, why the memory tensor can become
nontrivial, and why the combined stress tensor must be conserved on shell.

At the same time, the limitations are essential. This section does not claim
that the black-hole information paradox is solved. It does not derive a Page
curve, does not construct a unitary evaporation channel, does not prove
singularity resolution, and does not establish observational gravitational-wave
signatures. It also does not prove ghost freedom or perturbative stability
around a fully backreacted black-hole core. Each of these questions requires a
separate analysis after choosing a definite background, boundary conditions,
operator domain, activation function, memory potential, and matter sector.

The correct conclusion is therefore narrower and sharper. The present
formalism supplies a covariant, variationally consistent mechanism by which
black-hole curvature can activate a nonzero memory sector. The activated
sector is sourced by curvature history, carries a well-defined stress tensor,
is compatible with the Bianchi identity on shell, and can be organized as a
candidate core storage space for collapse and curvature-history data. In this
sense, the information-storage mechanism is not an external bookkeeping device:
it is encoded in the stress-energy-carrying memory fields of the black-hole
core. A complete theory of black-hole information recovery would require an
additional dynamical analysis of how this stored information is preserved,
processed, or released. Such an analysis is outside the scope of the present
foundational work and will be developed separately.

\section{Discussion and Conclusions}
\label{sec:discussion}

We have constructed a covariant auxiliary-field formulation for a
state-dependent infinite-derivative gravitational effective theory.  The
starting point was the observation that a nonlocality scale depending on the
curvature state of spacetime,
\begin{equation}
    M_* \longrightarrow M_{\rm eff}(C),
\end{equation}
cannot be inserted consistently unless the scalar \(C(x)\) is itself defined
as a covariant variational object.  A direct definition of \(C(x)\) as a
single-ray null-congruence curvature integral is geometrically suggestive but
not sufficiently robust for this purpose, because caustics, conjugate points,
branch dependence, and distributional metric variations obstruct a smooth
local stress-tensor construction.

The main result of the paper is the replacement of this curve-dependent
history variable by a local auxiliary curvature-history field \(C(x)\), with a
conjugate field \(\chi(x)\).  The field \(C\) obeys
\begin{equation}
    (\Box-m_c^2)C
    =
    -\ell_*\mathcal{K},
    \qquad
    \mathcal{K}
    =
    R_{\mu\nu\rho\sigma}R^{\mu\nu\rho\sigma},
    \label{eq:discussion_C_equation}
\end{equation}
and \(\chi\) enforces this equation at the level of the action.  With retarded
boundary data, \(C(x)\) admits the interpretation of a curvature-history
control field: it records invariant curvature information from the causal past
of \(x\).  In this sense, \(C\) is not an external regulator but a dynamical
state variable through which the nonlocal scale can respond to the geometry.

The physical mechanism enabled by this construction can be summarized as
follows.  Curvature generates \(C\) through Eq.~\eqref{eq:discussion_C_equation}.
The field \(C\) then controls the effective nonlocality scale
\(M_{\rm eff}(C)\), which enters the ordered form factor
\begin{equation}
    F_C
    =
    \exp\!\left[B(C)\Box_\Delta\right],
    \qquad
    B(C)=M_{\rm eff}^{-2}(C).
\end{equation}
Thus the strength and range of the infinite-derivative modification may depend
on the curvature history of the spacetime.  This provides a concrete mechanism
by which high-curvature regions, such as those arising during collapse,
black-hole core formation, or early-universe evolution, can dynamically alter
the scale at which nonlocal gravitational effects become important.

This mechanism is distinct from simply choosing a spacetime-dependent scale by
hand.  If \(M_{\rm eff}(C)\) is inserted without a variational equation for
\(C\), the modified Einstein equation need not be compatible with the
contracted Bianchi identity.  In the present formulation, the \(C\)-dependence
of the memory-sector action produces a source
\begin{equation}
    \Sigma_\Delta
    =
    \frac{1}{\sqrt{-g}}
    \frac{\delta S_\Delta}{\delta C}.
\end{equation}
The same source appears in the conjugate auxiliary equation,
\begin{equation}
    (\Box-m_c^2)\chi
    =
    -\Sigma_\Delta.
\end{equation}
Consequently, the memory sector and auxiliary history sector exchange
stress-energy locally.  On shell,
\begin{equation}
    \nabla^\mu T_{\mu\nu}^{(\Delta)}
    =
    \Sigma_\Delta\nabla_\nu C,
    \qquad
    \nabla^\mu T_{\mu\nu}^{(C)}
    =
    -\Sigma_\Delta\nabla_\nu C,
\end{equation}
and therefore
\begin{equation}
    \nabla^\mu
    \left(
        T_{\mu\nu}^{(\Delta)}
        +
        T_{\mu\nu}^{(C)}
    \right)
    =
    0.
    \label{eq:discussion_conservation}
\end{equation}
This cancellation is the central consistency result of the paper.  It shows
that a curvature-history-dependent nonlocal scale can be introduced without
violating the local conservation law required by diffeomorphism invariance.

A further technical point is the variation of the state-dependent nonlocal
operator.  Since \(C(x)\) is not constant, the multiplication operator
\(B(C)\) does not commute with \(\Box_\Delta\).  The form factor must therefore
be varied using Duhamel's formula rather than an ordinary scalar chain rule.
This fixes the kinetic contribution \(\Sigma_{\rm kin}\) to
\(\Sigma_\Delta\) and ensures that the source appearing in the \(\chi\)
equation is the same source required by the Noether identity.  The operator
ordering, adjoint structure, and Duhamel variation are therefore not merely
formal details; they are necessary for the conservation proof.

The representative black-hole application developed in
Sec.~\ref{sec:black_hole_core_application} shows how the formalism acquires a
concrete physical interpretation.  In the Schwarzschild reference geometry, the
Kretschmann scalar
\begin{equation}
    \mathcal{K}_{\rm Schw}(r)
    =
    \frac{48G_N^2M^2}{r^6}
\end{equation}
sources the curvature-history field.  In Ricci-flat black-hole regions this
source is equivalently the quadratic Weyl invariant,
\begin{equation}
    \mathcal{K}
    =
    W_{\alpha\beta\rho\sigma}W^{\alpha\beta\rho\sigma}.
\end{equation}
Thus the memory-history sector can respond to vacuum tidal curvature, not only
to local matter density.  A scalar activation domain can be defined by
\(\mathcal{K}\geq\mathcal{K}_{\rm crit}\) and \(C\geq C_{\rm crit}\).  With a
smooth activation function \(\Theta_C(C)\) and an activated potential, the core
branch satisfies
\begin{equation}
    \Delta_{\mu\nu}\Delta^{\mu\nu}
    \geq
    \frac{\Delta_*^2}{2}
    >0,
\end{equation}
so the memory tensor enters a nonvanishing phase in the activated region.

The same application also clarifies why the memory phase is physically active
rather than only formally defined.  The activated sector contributes through
\begin{equation}
    T_{\mu\nu}^{\rm mem}
    =
    T_{\mu\nu}^{(\Delta)}
    +
    T_{\mu\nu}^{(C)},
    \qquad
    \nabla^\mu T_{\mu\nu}^{\rm mem}=0
\end{equation}
on shell.  A representative stabilizing term
\(V_{\rm st}(C)=\rho_\Delta\Theta_C(C)\) gives a positive local timelike
contribution in the activated core,
\begin{equation}
    T_{\mu\nu}^{\rm st}n^\mu n^\nu
    =
    \rho_\Delta\Theta_C(C)
    \geq
    \frac{\rho_\Delta}{2}
    >0.
\end{equation}
The storage interpretation is then formulated in terms of an encoding map from
collapse or curvature histories to the gauge-quotient core memory data.  In
this limited sense, the information-storage mechanism is not an external
bookkeeping device: curvature history is encoded in fields that become
nonzero in the black-hole core and that contribute to the geometry through a
conserved stress tensor.

This black-hole application is intentionally limited.  It demonstrates that
the formalism has concrete physical content, but it does not claim to solve the
black-hole information problem.  It does not derive a Page curve, construct a
unitary evaporation channel, prove singularity resolution, or establish
observational gravitational-wave signatures.  It also does not prove ghost
freedom or perturbative stability around a fully backreacted black-hole core.
Those questions require a dedicated application paper in which the coupled
Einstein--memory equations, boundary conditions, stability, backreaction,
memory capacity, and possible information-release channels are analyzed in
detail.

The framework also clarifies how other physical applications should be
approached.  For gravitational collapse or compact-object models, one must
choose a specific function \(M_{\rm eff}(C)\), a potential \(V(\Delta,C)\),
and a coupling \(G(C)\), then solve the coupled system for a given symmetry
class and matter source.  The relevant questions are whether curvature
invariants remain finite, whether horizons form, whether the effective stress
tensor satisfies or violates particular energy conditions, and whether
perturbations remain stable.  The present paper does not answer all of these
questions, but it supplies the covariant field equations and conservation law
required before such questions can be posed consistently.

The same point applies to possible dark-sector applications.  The fields
\(C\), \(\chi\), and \(\Delta_{\mu\nu}\) contribute to the right-hand side of
the metric equation through
\begin{equation}
    T_{\mu\nu}^{\rm eff}
    =
    T_{\mu\nu}^{(\Delta)}
    +
    T_{\mu\nu}^{(C)}.
\end{equation}
This effective geometric stress tensor is conserved on shell and may, in
specific backgrounds, behave as an additional gravitational source.  Therefore
the formalism provides a possible route to studying whether curvature-history
degrees of freedom can generate effective anisotropic stresses or mass profiles
relevant to dark-sector phenomenology.  However, no dark-matter profile or
galactic-scale phenomenology is derived here.  Such claims require explicit
solutions, parameter constraints, and comparison with observational data.

The construction is also naturally connected to the broader modified-gravity
literature on compact objects, collapse, stability, and curvature corrections.
Recent studies in \(f(R,T)\), Palatini \(F(R)\), \(f(R)\), and \(f(R,G)\)
gravity have emphasized that modified curvature terms can affect instability
conditions, effective forces, structure scalars, and compact-object dynamics
\cite{Bhatti2020Stability,Asad2024Palatini,Yousaf2025AxialCollapse,Rehman2025Complexity}.
The present work is different in its specific use of an infinite-derivative
state-dependent form factor, but it shares with those studies the broader goal
of understanding how additional geometric structure can influence strongly
gravitating systems.

Several limitations should be emphasized.  First, the conservation law derived
in Eq.~\eqref{eq:discussion_conservation} is an on-shell identity; it does not
by itself prove existence, uniqueness, or global well-posedness of the coupled
system.  Second, the use of an entire form factor does not automatically prove
ghost freedom around arbitrary backgrounds when the scale is state dependent.
A spectral analysis must be performed after choosing a background and a
specific form of \(M_{\rm eff}(C)\).  Third, although \(C\) admits a retarded
history interpretation, the causal properties of the full infinite-derivative
system require a separate analysis of the chosen operator domain and boundary
conditions.  Fourth, the present paper does not derive singularity resolution,
a Page curve, gravitational-wave signatures, or dark-matter phenomenology.

The natural next step is therefore to apply the formalism to explicit
backgrounds.  For black holes, the immediate application is a full analysis of
curvature-triggered core memory: one should solve the coupled equations in a
black-hole or collapse background, study the backreaction of
\(T_{\mu\nu}^{\rm mem}\), determine the stability of the activated phase,
construct the core memory phase space, and test whether the encoding map is
nondegenerate for physically relevant collapse histories.  For compact objects,
one should study static and dynamical spherically symmetric ans\"{a}tze, compute
curvature invariants, analyze the effective radial and tangential pressures,
and test linear stability.  For cosmology, one should derive the homogeneous
and perturbed equations for \(C\), \(\chi\), and \(\Delta_{\mu\nu}\), and
determine whether the curvature-history sector can produce controlled
deviations from standard early- or late-time evolution.  For the quantum
consistency of the model, one should analyze the pole structure of the
perturbative propagator around backgrounds with nonconstant \(C\).

In conclusion, this paper establishes a covariant action-level mechanism for
state-dependent infinite-derivative gravity.  The auxiliary curvature-history
field \(C\) provides a smooth replacement for a curve-dependent null-memory
variable, the conjugate field \(\chi\) incorporates the response of the memory
sector, and Duhamel's formula gives the correct variation of the
state-dependent nonlocal operator.  The resulting on-shell Noether identity
shows that the combined memory-history stress tensor is covariantly conserved.
The representative black-hole application shows why this consistency result is
physically useful: black-hole curvature can source the history field, activate
a nonzero memory phase, and produce a stress-energy-carrying core sector that
can encode curvature-history data.  A complete black-hole information-storage
or information-recovery theory requires a separate dynamical analysis, but the
present work supplies the covariant foundation on which such an application can
be built.


\section*{Acknowledgments}

The author thanks the editor and the anonymous reviewers for their careful
reading of the manuscript.  The author is especially grateful for the detailed
and constructive suggestions that helped improve the motivation, organization,
operator-ordering conventions, and presentation of the revised work.
\appendix

\section{Metric Variation of the Kretschmann-History Coupling}
\label{app:kretschmann_variation}

This appendix derives the metric variation of the curvature-squared coupling
that appears in the auxiliary curvature-history sector.  The result is used in
Sec.~\ref{sec:metric_variation} for the stress-energy tensor associated with
\(\ell_*\chi\mathcal{K}\).

Let \(f\) be a smooth scalar field held fixed under metric variation.  In the
application to the auxiliary curvature-history sector one sets
\begin{equation}
    f=\chi .
\end{equation}
Define
\begin{equation}
    I_{\mathcal{K}}[g;f]
    =
    \int_{\mathcal{M}} d^4x\,\sqrt{-g}\,
    f\,\mathcal{K},
    \qquad
    \mathcal{K}
    =
    R_{\alpha\beta\rho\sigma}R^{\alpha\beta\rho\sigma}.
    \label{eq:IK_def_appA}
\end{equation}
The corresponding variational tensor is
\begin{equation}
    \Theta_{\mu\nu}[f]
    \equiv
    -
    \frac{2}{\sqrt{-g}}
    \frac{\delta I_{\mathcal{K}}[g;f]}{\delta g^{\mu\nu}} .
    \label{eq:Theta_def_appA}
\end{equation}

We vary \(I_{\mathcal{K}}\) with respect to the inverse metric
\(g^{\mu\nu}\).  The volume element varies as
\begin{equation}
    \delta\sqrt{-g}
    =
    -
    \frac12\sqrt{-g}\,
    g_{\mu\nu}\delta g^{\mu\nu}.
    \label{eq:volume_variation_appA}
\end{equation}
Since \(f\) is held fixed,
\begin{equation}
    \delta I_{\mathcal{K}}
    =
    \int_{\mathcal{M}} d^4x\,\sqrt{-g}
    \left[
        -
        \frac12 g_{\mu\nu}f\mathcal{K}\,\delta g^{\mu\nu}
        +
        f\,\delta\mathcal{K}
    \right].
    \label{eq:delta_IK_start_appA}
\end{equation}

It is useful to write the Kretschmann invariant in the form
\begin{equation}
    \mathcal{K}
    =
    R^\alpha{}_{\beta\rho\sigma}
    R_\alpha{}^{\beta\rho\sigma}.
    \label{eq:K_mixed_form_appA}
\end{equation}
With this choice, the variation separates cleanly into an algebraic metric
contraction part and a Palatini curvature-variation part:
\begin{equation}
    \delta\mathcal{K}
    =
    2R_{\mu\alpha\beta\gamma}
    R_{\nu}{}^{\alpha\beta\gamma}
    \delta g^{\mu\nu}
    +
    2R_\alpha{}^{\beta\rho\sigma}
    \delta R^\alpha{}_{\beta\rho\sigma}.
    \label{eq:delta_K_split_appA}
\end{equation}
This form avoids double counting algebraic metric variations from lowering the
first index of the Riemann tensor.

With the convention
\begin{equation}
    [\nabla_\mu,\nabla_\nu]V^\rho
    =
    R^\rho{}_{\sigma\mu\nu}V^\sigma,
\end{equation}
the Palatini identity gives
\begin{equation}
    \delta R^\alpha{}_{\beta\rho\sigma}
    =
    \nabla_\rho\delta\Gamma^\alpha_{\sigma\beta}
    -
    \nabla_\sigma\delta\Gamma^\alpha_{\rho\beta},
    \label{eq:palatini_identity_appA}
\end{equation}
where
\begin{equation}
    \delta\Gamma^\alpha_{\rho\beta}
    =
    \frac12 g^{\alpha\lambda}
    \left(
        \nabla_\rho\delta g_{\beta\lambda}
        +
        \nabla_\beta\delta g_{\rho\lambda}
        -
        \nabla_\lambda\delta g_{\rho\beta}
    \right).
    \label{eq:delta_connection_appA}
\end{equation}
Using the antisymmetry of the Riemann tensor in its last two indices,
\begin{align}
    2fR_\alpha{}^{\beta\rho\sigma}
    \delta R^\alpha{}_{\beta\rho\sigma}
    &=
    2fR_\alpha{}^{\beta\rho\sigma}
    \left(
        \nabla_\rho\delta\Gamma^\alpha_{\sigma\beta}
        -
        \nabla_\sigma\delta\Gamma^\alpha_{\rho\beta}
    \right)
    \nonumber \\
    &=
    4fR_\lambda{}^{\beta\rho\sigma}
    \nabla_\rho\delta\Gamma^\lambda_{\sigma\beta}.
    \label{eq:derivative_part_start_appA}
\end{align}

The derivative part of the variation is therefore
\begin{equation}
    \delta I_{\mathcal{K}}\big|_{\rm der}
    =
    4
    \int_{\mathcal{M}} d^4x\,\sqrt{-g}\,
    fR_\lambda{}^{\beta\rho\sigma}
    \nabla_\rho\delta\Gamma^\lambda_{\sigma\beta}.
    \label{eq:delta_IK_der_start_appA}
\end{equation}
Integrating by parts once gives
\begin{equation}
    \delta I_{\mathcal{K}}\big|_{\rm der}
    =
    -
    4
    \int_{\mathcal{M}} d^4x\,\sqrt{-g}\,
    \nabla_\rho
    \left(
        fR_\lambda{}^{\beta\rho\sigma}
    \right)
    \delta\Gamma^\lambda_{\sigma\beta},
    \label{eq:first_ibp_appA}
\end{equation}
up to a boundary contribution.  Substituting
Eq.~\eqref{eq:delta_connection_appA}, using
\begin{equation}
    \delta g_{\mu\nu}
    =
    -
    g_{\mu\alpha}g_{\nu\beta}\delta g^{\alpha\beta},
    \label{eq:delta_lower_metric_appA}
\end{equation}
and integrating by parts a second time yields a contribution of the form
\begin{equation}
    \delta I_{\mathcal{K}}\big|_{\rm der}
    =
    \int_{\mathcal{M}} d^4x\,\sqrt{-g}\,
    \mathcal{D}_{\mu\nu}[f]\,
    \delta g^{\mu\nu}.
    \label{eq:delta_IK_der_final_form_appA}
\end{equation}
Because the independent metric variation \(\delta g^{\mu\nu}\) is symmetric,
only the symmetric part of the coefficient contributes.  The derivative
coefficient may therefore be written in the explicitly symmetric form
\begin{equation}
    \mathcal{D}_{\mu\nu}[f]
    =
    2\nabla^\alpha\nabla^\beta
    \left[
        f
        \left(
            R_{\mu\alpha\nu\beta}
            +
            R_{\nu\alpha\mu\beta}
        \right)
    \right].
    \label{eq:D_symmetric_appA}
\end{equation}
Equivalently,
\begin{equation}
    \mathcal{D}_{\mu\nu}[f]
    =
    4\nabla^\alpha\nabla^\beta
    \left(
        f R_{\mu\alpha\nu\beta}
    \right)
\end{equation}
when the expression is understood under contraction with the symmetric
variation \(\delta g^{\mu\nu}\).  The form in
Eq.~\eqref{eq:D_symmetric_appA} is preferable because the symmetry
\(\mathcal{D}_{\mu\nu}=\mathcal{D}_{\nu\mu}\) is manifest.

Combining the volume variation, the algebraic part of
Eq.~\eqref{eq:delta_K_split_appA}, and the derivative contribution
Eq.~\eqref{eq:D_symmetric_appA}, the full metric variation can be written as
\begin{equation}
    \delta I_{\mathcal{K}}
    =
    \int_{\mathcal{M}} d^4x\,\sqrt{-g}\,
    \mathcal{H}_{\mu\nu}[f]\,
    \delta g^{\mu\nu},
    \label{eq:delta_IK_final_appA}
\end{equation}
where
\begin{equation}
    \mathcal{H}_{\mu\nu}[f]
    =
    -
    \frac12 g_{\mu\nu}f\mathcal{K}
    +
    2fR_{\mu\alpha\beta\gamma}
    R_\nu{}^{\alpha\beta\gamma}
    +
    2\nabla^\alpha\nabla^\beta
    \left[
        f
        \left(
            R_{\mu\alpha\nu\beta}
            +
            R_{\nu\alpha\mu\beta}
        \right)
    \right].
    \label{eq:H_tensor_appA}
\end{equation}
This tensor is explicitly symmetric in \(\mu\) and \(\nu\), as required for a
metric variational derivative.

Using the definition
\eqref{eq:Theta_def_appA}, one obtains
\begin{equation}
    \boxed{
    \Theta_{\mu\nu}[f]
    =
    g_{\mu\nu}f\mathcal{K}
    -
    4fR_{\mu\alpha\beta\gamma}
    R_\nu{}^{\alpha\beta\gamma}
    -
    4\nabla^\alpha\nabla^\beta
    \left[
        f
        \left(
            R_{\mu\alpha\nu\beta}
            +
            R_{\nu\alpha\mu\beta}
        \right)
    \right]
    } .
    \label{eq:Theta_result_appA}
\end{equation}

For the auxiliary curvature-history action,
\begin{equation}
    S_C
    \supset
    \ell_*
    \int_{\mathcal{M}} d^4x\,\sqrt{-g}\,
    \chi\mathcal{K},
\end{equation}
one sets \(f=\chi\).  Hence the Kretschmann-history contribution to the
auxiliary stress tensor is
\begin{equation}
    \ell_*\Theta_{\mu\nu}[\chi]
    =
    \ell_*
    \left\{
        g_{\mu\nu}\chi\mathcal{K}
        -
        4\chi R_{\mu\alpha\beta\gamma}
        R_\nu{}^{\alpha\beta\gamma}
        -
        4\nabla^\alpha\nabla^\beta
        \left[
            \chi
            \left(
                R_{\mu\alpha\nu\beta}
                +
                R_{\nu\alpha\mu\beta}
            \right)
        \right]
    \right\}.
    \label{eq:Theta_chi_result_appA}
\end{equation}
This is the curvature-squared contribution appearing in the auxiliary
stress-energy tensor.

\section{Mass Dimensions and Sign Conventions}
\label{app:dimensions_signs}

This appendix collects the mass-dimensional assignments and sign conventions
used throughout the paper.  These conventions are stated here in one place to
avoid repetition in the main text.

\subsection{Mass dimensions}
\label{subsec:mass_dimensions_appB}

We work in natural units,
\begin{equation}
    c=\hbar=1.
\end{equation}
The action is dimensionless:
\begin{equation}
    [S]=0.
    \label{eq:action_dimension_appB}
\end{equation}
The coordinate and derivative dimensions are
\begin{equation}
    [x^\mu]=-1,
    \qquad
    [\partial_\mu]=1,
    \qquad
    [\nabla_\mu]=1.
    \label{eq:coordinate_derivative_dimensions_appB}
\end{equation}
The metric is taken to be dimensionless:
\begin{equation}
    [g_{\mu\nu}]=[g^{\mu\nu}]=0,
    \qquad
    [\sqrt{-g}]=0.
    \label{eq:metric_dimensions_appB}
\end{equation}
Since
\begin{equation}
    [d^4x]=-4,
\end{equation}
the scalar Lagrangian density inside
\(\int d^4x\sqrt{-g}\,\mathcal{L}\) must have dimension
\begin{equation}
    [\mathcal{L}]=4.
    \label{eq:L_dimension_appB}
\end{equation}

With the curvature convention
\begin{equation}
    [\nabla_\mu,\nabla_\nu]V^\rho
    =
    R^\rho{}_{\sigma\mu\nu}V^\sigma,
\end{equation}
one has
\begin{equation}
    [\Gamma^\rho_{\mu\nu}]=1,
    \qquad
    [R^\rho{}_{\sigma\mu\nu}]=2,
    \qquad
    [R_{\mu\nu}]=2,
    \qquad
    [R]=2.
    \label{eq:curvature_dimensions_appB}
\end{equation}
The Kretschmann invariant therefore has dimension
\begin{equation}
    [\mathcal{K}]
    =
    [R_{\mu\nu\rho\sigma}R^{\mu\nu\rho\sigma}]
    =
    4.
    \label{eq:K_dimension_appB}
\end{equation}

The matter scalar, memory tensor, and auxiliary scalars are assigned
\begin{equation}
    [\Phi]=1,
    \qquad
    [\Delta_{\mu\nu}]=1,
    \qquad
    [C]=1,
    \qquad
    [\chi]=1.
    \label{eq:field_dimensions_appB}
\end{equation}
The mass parameters and the curvature-history length scale have dimensions
\begin{equation}
    [m_c]=1,
    \qquad
    [M_*]=1,
    \qquad
    [C_*]=1,
    \qquad
    [M_{\rm eff}]=1,
    \qquad
    [\ell_*]=-1.
    \label{eq:scale_dimensions_appB}
\end{equation}
Thus
\begin{equation}
    [M_{\rm eff}^2]=2,
    \qquad
    [B(C)]=[M_{\rm eff}^{-2}(C)]=-2.
    \label{eq:B_dimension_appB}
\end{equation}

These assignments make each term in the auxiliary curvature-history Lagrangian
dimension four:
\begin{equation}
    [\nabla_\mu\chi\nabla^\mu C]=4,
    \qquad
    [m_c^2\chi C]=4,
    \qquad
    [\ell_*\chi\mathcal{K}]=4.
    \label{eq:LC_dimension_check_appB}
\end{equation}
They also make the memory kinetic term dimension four:
\begin{equation}
    [\Delta_{\mu\nu}\Box_\Delta\Delta^{\mu\nu}]
    =
    1+2+1
    =
    4.
    \label{eq:Delta_kin_dimension_appB}
\end{equation}
Since
\begin{equation}
    [B(C)\Box_\Delta]=0,
\end{equation}
the ordered exponent
\begin{equation}
    F_C
    =
    \exp[B(C)\Box_\Delta]
\end{equation}
is dimensionless.

For convenience, the assignments are summarized in
Table~\ref{tab:dimension_summary_appB}.

\begin{table}[t]
\centering
\begin{tabular}{c c}
\toprule
Quantity & Mass dimension \\
\midrule
\(x^\mu\) & \(-1\) \\
\(\partial_\mu,\nabla_\mu\) & \(1\) \\
\(g_{\mu\nu},g^{\mu\nu},\sqrt{-g}\) & \(0\) \\
\(\Gamma^\rho_{\mu\nu}\) & \(1\) \\
\(R^\rho{}_{\sigma\mu\nu},R_{\mu\nu},R\) & \(2\) \\
\(\mathcal{K}\) & \(4\) \\
\(\Phi\) & \(1\) \\
\(\Delta_{\mu\nu}\) & \(1\) \\
\(C,\chi\) & \(1\) \\
\(m_c,M_*,C_*,M_{\rm eff}\) & \(1\) \\
\(B(C)=M_{\rm eff}^{-2}(C)\) & \(-2\) \\
\(\ell_*\) & \(-1\) \\
\bottomrule
\end{tabular}
\caption{Mass-dimensional assignments used in the manuscript.}
\label{tab:dimension_summary_appB}
\end{table}

\subsection{Curvature and d'Alembertian conventions}
\label{subsec:curvature_signs_appB}

The Riemann tensor convention is
\begin{equation}
    [\nabla_\mu,\nabla_\nu]V^\rho
    =
    R^\rho{}_{\sigma\mu\nu}V^\sigma.
    \label{eq:riemann_convention_appB}
\end{equation}
The Ricci tensor and Ricci scalar are
\begin{equation}
    R_{\mu\nu}=R^\rho{}_{\mu\rho\nu},
    \qquad
    R=g^{\mu\nu}R_{\mu\nu}.
    \label{eq:ricci_convention_appB}
\end{equation}
The Einstein tensor is
\begin{equation}
    G_{\mu\nu}
    =
    R_{\mu\nu}
    -
    \frac12 g_{\mu\nu}R,
    \label{eq:Einstein_tensor_appB}
\end{equation}
and satisfies
\begin{equation}
    \nabla^\mu G_{\mu\nu}=0.
    \label{eq:Bianchi_appB}
\end{equation}
The scalar d'Alembertian is
\begin{equation}
    \Box
    =
    g^{\mu\nu}\nabla_\mu\nabla_\nu.
    \label{eq:box_scalar_appB}
\end{equation}
For a symmetric tensor \(X_{\mu\nu}\), the rough tensor d'Alembertian is
\begin{equation}
    (\Box_\Delta X)_{\mu\nu}
    =
    g^{\rho\sigma}\nabla_\rho\nabla_\sigma X_{\mu\nu}.
    \label{eq:box_tensor_appB}
\end{equation}

\subsection{Auxiliary-sector sign conventions}
\label{subsec:auxiliary_signs_appB}

The auxiliary curvature-history action is
\begin{equation}
    S_C
    =
    \int_{\mathcal{M}}d^4x\sqrt{-g}
    \left[
        -
        \nabla_\mu\chi\nabla^\mu C
        -
        m_c^2\chi C
        +
        \ell_*\chi\mathcal{K}
    \right].
    \label{eq:SC_appB}
\end{equation}
With this convention, variation with respect to \(\chi\) gives
\begin{equation}
    \delta_\chi S_C
    =
    \int_{\mathcal{M}}d^4x\sqrt{-g}\,
    \delta\chi
    \left[
        \Box C
        -
        m_c^2 C
        +
        \ell_*\mathcal{K}
    \right].
    \label{eq:delta_chi_SC_appB}
\end{equation}
Therefore the curvature-history equation is
\begin{equation}
    (\Box-m_c^2)C
    =
    -\ell_*\mathcal{K}.
    \label{eq:C_equation_appB}
\end{equation}

Variation of \(S_C\) with respect to \(C\) gives
\begin{equation}
    \delta_C S_C
    =
    \int_{\mathcal{M}}d^4x\sqrt{-g}\,
    \delta C
    \left[
        \Box\chi
        -
        m_c^2\chi
    \right].
    \label{eq:delta_C_SC_appB}
\end{equation}
The memory-sector source \(\Sigma_\Delta\) is defined by
\begin{equation}
    \delta_C S_\Delta
    =
    \int_{\mathcal{M}}d^4x\sqrt{-g}\,
    \Sigma_\Delta\,\delta C .
    \label{eq:Sigma_definition_appB}
\end{equation}
Thus the full \(C\)-variation of \(S_C+S_\Delta\) gives
\begin{equation}
    \Box\chi
    -
    m_c^2\chi
    +
    \Sigma_\Delta
    =
    0,
    \label{eq:chi_equation_raw_appB}
\end{equation}
or
\begin{equation}
    (\Box-m_c^2)\chi
    =
    -\Sigma_\Delta.
    \label{eq:chi_equation_appB}
\end{equation}

With the same source convention, the decomposition of
\(\Sigma_\Delta\) is
\begin{equation}
    \Sigma_\Delta
    =
    \Sigma_{\rm kin}
    -
    \frac{\partial V}{\partial C}
    +
    \eta G'(C)\Delta_{\mu\nu}R^{\mu\nu}.
    \label{eq:Sigma_decomposition_appB}
\end{equation}
Here
\begin{equation}
    \delta_C S_\Delta^{\rm kin}
    =
    \int_{\mathcal{M}}d^4x\sqrt{-g}\,
    \Sigma_{\rm kin}\,\delta C.
    \label{eq:Sigma_kin_definition_appB}
\end{equation}

\subsection{Stress-tensor sign conventions}
\label{subsec:stress_tensor_signs_appB}

All stress-energy tensors are defined by variation with respect to the inverse
metric:
\begin{equation}
    T_{\mu\nu}^{(X)}
    =
    -
    \frac{2}{\sqrt{-g}}
    \frac{\delta S_X}{\delta g^{\mu\nu}}.
    \label{eq:stress_tensor_def_appB}
\end{equation}
Equivalently,
\begin{equation}
    \delta_g S_X
    =
    -
    \frac12
    \int_{\mathcal{M}}d^4x\sqrt{-g}\,
    T_{\mu\nu}^{(X)}\delta g^{\mu\nu}.
    \label{eq:stress_tensor_variation_appB}
\end{equation}
The Einstein--Hilbert variation is written as
\begin{equation}
    \delta_g S_{\rm EH}
    =
    \frac{1}{16\pi G_N}
    \int_{\mathcal{M}}d^4x\sqrt{-g}\,
    G_{\mu\nu}\delta g^{\mu\nu},
    \label{eq:EH_variation_appB}
\end{equation}
up to boundary terms.  Therefore the metric equation is
\begin{equation}
    G_{\mu\nu}
    =
    8\pi G_N
    \left(
        T_{\mu\nu}^{\rm mat}
        +
        T_{\mu\nu}^{(\Delta)}
        +
        T_{\mu\nu}^{(C)}
    \right).
    \label{eq:metric_equation_appB}
\end{equation}

The Kretschmann-history contribution is defined by
\begin{equation}
    \Theta_{\mu\nu}[f]
    =
    -
    \frac{2}{\sqrt{-g}}
    \frac{\delta}{\delta g^{\mu\nu}}
    \int_{\mathcal{M}}d^4x\sqrt{-g}\,
    f\mathcal{K}.
    \label{eq:Theta_def_appB}
\end{equation}
With the conventions of Appendix~\ref{app:kretschmann_variation},
\begin{equation}
    \Theta_{\mu\nu}[f]
    =
    g_{\mu\nu}f\mathcal{K}
    -
    4fR_{\mu\alpha\beta\gamma}
    R_\nu{}^{\alpha\beta\gamma}
    -
    4\nabla^\alpha\nabla^\beta
    \left[
        f
        \left(
            R_{\mu\alpha\nu\beta}
            +
            R_{\nu\alpha\mu\beta}
        \right)
    \right].
    \label{eq:Theta_result_appB}
\end{equation}
The auxiliary curvature-history stress tensor is therefore
\begin{align}
    T_{\mu\nu}^{(C)}
    &=
    2\nabla_{(\mu}\chi\nabla_{\nu)}C
    -
    g_{\mu\nu}\nabla_\rho\chi\nabla^\rho C
    -
    g_{\mu\nu}m_c^2\chi C
    \nonumber \\
    &\quad
    +
    \ell_*\Theta_{\mu\nu}[\chi].
    \label{eq:TC_appB}
\end{align}

\subsection{Noether-exchange sign convention}
\label{subsec:noether_exchange_signs_appB}

The active infinitesimal diffeomorphism variation used in the conservation
proof is
\begin{equation}
    \delta_\xi=\mathcal{L}_\xi.
    \label{eq:lie_convention_appB}
\end{equation}
Thus
\begin{equation}
    \delta_\xi g^{\mu\nu}
    =
    -2\nabla^{(\mu}\xi^{\nu)},
    \qquad
    \delta_\xi C
    =
    \xi^\rho\nabla_\rho C.
    \label{eq:lie_variations_appB}
\end{equation}
Using
\begin{equation}
    \delta_C S_\Delta
    =
    \int d^4x\sqrt{-g}\,
    \Sigma_\Delta\delta C,
\end{equation}
the memory-sector Noether identity gives, on the memory-field equation,
\begin{equation}
    \nabla^\mu T_{\mu\nu}^{(\Delta)}
    =
    \Sigma_\Delta\nabla_\nu C.
    \label{eq:TDelta_divergence_appB}
\end{equation}
The auxiliary-sector identity gives, on the full \(C\) and \(\chi\) equations,
\begin{equation}
    \nabla^\mu T_{\mu\nu}^{(C)}
    =
    -
    \Sigma_\Delta\nabla_\nu C.
    \label{eq:TC_divergence_appB}
\end{equation}
Adding Eqs.~\eqref{eq:TDelta_divergence_appB} and
\eqref{eq:TC_divergence_appB} yields
\begin{equation}
    \nabla^\mu
    \left(
        T_{\mu\nu}^{(\Delta)}
        +
        T_{\mu\nu}^{(C)}
    \right)
    =
    0.
    \label{eq:combined_conservation_appB}
\end{equation}

If the source is instead defined with the opposite sign,
\begin{equation}
    \delta_C S_\Delta
    =
    -
    \int d^4x\sqrt{-g}\,
    \widetilde{\Sigma}_\Delta\delta C,
\end{equation}
then the two exchange terms in
Eqs.~\eqref{eq:TDelta_divergence_appB} and
\eqref{eq:TC_divergence_appB} both reverse sign.  The combined conservation law
is unchanged.

\subsection{Operator-ordering convention}
\label{subsec:operator_ordering_signs_appB}

The ordered nonlocal operator used in the manuscript is
\begin{equation}
    \mathcal{A}_C
    =
    B(C)\Box_\Delta,
    \qquad
    B(C)=M_{\rm eff}^{-2}(C),
    \label{eq:A_ordering_appB}
\end{equation}
and
\begin{equation}
    F_C
    =
    \exp(\mathcal{A}_C)
    =
    \exp[B(C)\Box_\Delta].
    \label{eq:F_ordering_appB}
\end{equation}
All products are composed from right to left:
\begin{equation}
    (\mathcal{A}_C X)_{\mu\nu}
    =
    B(C)(\Box_\Delta X)_{\mu\nu}.
    \label{eq:A_action_appB}
\end{equation}
This convention fixes the \(C\)-variation
\begin{equation}
    \delta_C\mathcal{A}_C
    =
    B'(C)\delta C\,\Box_\Delta.
    \label{eq:delta_A_appB}
\end{equation}
The corresponding noncommuting variation of \(F_C\) is given by Duhamel's
formula in Appendix~\ref{app:operator_identities}.

\section{Operator Ordering, Formal Adjoints, and Duhamel Identities}
\label{app:operator_identities}

This appendix collects the operator conventions and identities used in the
variation of the state-dependent nonlocal form factor.  The main text uses
these results in Secs.~\ref{sec:covariant_auxiliary_action},
\ref{sec:duhamel_variation}, and \ref{sec:memory_field_equation}.

\subsection{Ordered nonlocal operator}
\label{subsec:ordered_operator_appC}

The state-dependent inverse mass scale is denoted by
\begin{equation}
    B(C)
    =
    M_{\rm eff}^{-2}(C).
    \label{eq:B_def_appC}
\end{equation}
It acts by scalar multiplication.  The tensor d'Alembertian acting on a
symmetric rank-two tensor \(X_{\mu\nu}\) is
\begin{equation}
    (\Box_\Delta X)_{\mu\nu}
    =
    g^{\rho\sigma}\nabla_\rho\nabla_\sigma X_{\mu\nu}.
    \label{eq:tensor_box_appC}
\end{equation}
The ordered exponent used in the manuscript is
\begin{equation}
    \mathcal{A}_C
    =
    B(C)\Box_\Delta .
    \label{eq:A_def_appC}
\end{equation}
Operator products are composed from right to left.  Therefore
\begin{equation}
    (\mathcal{A}_C X)_{\mu\nu}
    =
    B(C)(\Box_\Delta X)_{\mu\nu},
    \label{eq:A_action_appC}
\end{equation}
and
\begin{equation}
    (\mathcal{A}_C^2 X)_{\mu\nu}
    =
    B(C)
    \Box_\Delta
    \left[
        B(C)(\Box_\Delta X)_{\mu\nu}
    \right].
    \label{eq:A_squared_appC}
\end{equation}
The form factor is defined by the analytic series
\begin{equation}
    F_C
    =
    \exp(\mathcal{A}_C)
    =
    \sum_{n=0}^{\infty}
    \frac{1}{n!}\mathcal{A}_C^n.
    \label{eq:F_series_appC}
\end{equation}
Thus
\begin{equation}
    F_C
    =
    \exp[B(C)\Box_\Delta]
    \label{eq:F_ordered_appC}
\end{equation}
always denotes the ordered operator in Eq.~\eqref{eq:A_def_appC}.  A different
ordering, such as \(\Box_\Delta B(C)\), would define a different model.

\subsection{Commutator with scalar multiplication}
\label{subsec:commutator_appC}

Because \(C\) is spacetime dependent, scalar multiplication by \(B(C)\) does
not generally commute with \(\Box_\Delta\).  Acting on a symmetric tensor
\(X_{\mu\nu}\),
\begin{align}
    [\Box_\Delta,B]X_{\mu\nu}
    &=
    \Box_\Delta(BX_{\mu\nu})
    -
    B\Box_\Delta X_{\mu\nu}
    \nonumber \\
    &=
    (\Box B)X_{\mu\nu}
    +
    2(\nabla^\rho B)\nabla_\rho X_{\mu\nu}.
    \label{eq:box_B_commutator_appC}
\end{align}
Therefore
\begin{equation}
    [B(C),\Box_\Delta]\neq 0
    \label{eq:noncommuting_appC}
\end{equation}
unless \(B(C)\) is constant or the field being acted on obeys special
conditions.  This is the reason the variation of \(F_C\) must be treated by
Duhamel's formula.

\subsection{Formal adjoints}
\label{subsec:formal_adjoints_appC}

For smooth symmetric tensors satisfying the boundary conditions stated in
Sec.~\ref{subsec:regularity_assumptions_section2}, define the bilinear pairing
\begin{equation}
    \langle X,Y\rangle
    =
    \int_{\mathcal{M}} d^4x\,\sqrt{-g}\,
    X_{\mu\nu}Y^{\mu\nu}.
    \label{eq:pairing_appC}
\end{equation}
This is used only as a formal integration-by-parts pairing.  In Lorentzian
signature it is not a positive definite inner product.

With this pairing,
\begin{equation}
    \langle X,\Box_\Delta Y\rangle
    =
    \langle \Box_\Delta X,Y\rangle ,
    \label{eq:box_self_adjoint_appC}
\end{equation}
up to boundary terms.  Scalar multiplication by \(B(C)\) is formally
self-adjoint:
\begin{equation}
    \langle X,BY\rangle
    =
    \langle BX,Y\rangle .
    \label{eq:B_self_adjoint_appC}
\end{equation}
Consequently, the formal adjoint of
\(\mathcal{A}_C=B(C)\Box_\Delta\) is
\begin{equation}
    \mathcal{A}_C^\dagger
    =
    \Box_\Delta B(C),
    \label{eq:A_adjoint_appC}
\end{equation}
meaning
\begin{equation}
    \langle X,\mathcal{A}_C Y\rangle
    =
    \langle \mathcal{A}_C^\dagger X,Y\rangle .
    \label{eq:A_adjoint_definition_appC}
\end{equation}
Explicitly,
\begin{equation}
    (\mathcal{A}_C^\dagger X)_{\mu\nu}
    =
    \Box_\Delta\!\left[B(C)X_{\mu\nu}\right].
    \label{eq:A_adjoint_action_appC}
\end{equation}
Since \(B(C)\) and \(\Box_\Delta\) do not commute in general,
\begin{equation}
    \mathcal{A}_C^\dagger\neq \mathcal{A}_C.
\end{equation}
The adjoint of the form factor is
\begin{equation}
    F_C^\dagger
    =
    \exp(\mathcal{A}_C^\dagger).
    \label{eq:F_adjoint_appC}
\end{equation}

\subsection{Duhamel identity}
\label{subsec:duhamel_identity_appC}

For a differentiable one-parameter family of operators
\(\mathcal{A}(\epsilon)\), the first variation of the exponential is
\begin{equation}
    \delta e^{\mathcal{A}}
    =
    \int_0^1
    e^{s\mathcal{A}}
    (\delta\mathcal{A})
    e^{(1-s)\mathcal{A}}
    ds .
    \label{eq:duhamel_identity_appC}
\end{equation}
This is Duhamel's formula.  It follows by differentiating
\(e^{\mathcal{A}+\epsilon\delta\mathcal{A}}\) at \(\epsilon=0\), or
equivalently by expanding both exponentials in a power series and collecting
all possible insertions of \(\delta\mathcal{A}\).

For the operator used in this paper,
\begin{equation}
    \mathcal{A}_C
    =
    B(C)\Box_\Delta,
\end{equation}
the variation with respect to \(C\), at fixed metric and fixed tensor field, is
\begin{equation}
    \delta_C\mathcal{A}_C
    =
    B'(C)\delta C\,\Box_\Delta.
    \label{eq:delta_A_C_appC}
\end{equation}
Thus
\begin{equation}
    \delta_C F_C
    =
    \int_0^1
    e^{s\mathcal{A}_C}
    \left[
        B'(C)\delta C\,\Box_\Delta
    \right]
    e^{(1-s)\mathcal{A}_C}
    ds.
    \label{eq:delta_F_C_appC}
\end{equation}
This identity is used in Sec.~\ref{sec:duhamel_variation} to define the
nonlocal kinetic contribution \(\Sigma_{\rm kin}\) to the source
\(\Sigma_\Delta\).

When
\begin{equation}
    [\mathcal{A}_C,\delta_C\mathcal{A}_C]=0,
\end{equation}
Duhamel's formula reduces to the ordinary scalar rule
\begin{equation}
    \delta_C F_C
    =
    F_C\,\delta_C\mathcal{A}_C.
    \label{eq:commuting_limit_variation_appC}
\end{equation}
This special case occurs, for example, when \(C\) and \(\delta C\) are constant
on the region considered.  The general spacetime-dependent case requires
Eq.~\eqref{eq:delta_F_C_appC}.

\subsection{Adjoint form of the kinetic variation}
\label{subsec:kinetic_variation_identity_appC}

Let
\begin{equation}
    Y_{\mu\nu}
    =
    (\Box_\Delta\Delta)_{\mu\nu}.
    \label{eq:Y_def_appC}
\end{equation}
The nonlocal kinetic term is
\begin{equation}
    S_\Delta^{\rm kin}
    =
    -
    \frac12
    \langle \Delta,F_CY\rangle .
    \label{eq:Skin_pairing_appC}
\end{equation}
At fixed metric and fixed \(\Delta_{\mu\nu}\),
\begin{equation}
    \delta_C S_\Delta^{\rm kin}
    =
    -
    \frac12
    \langle \Delta,(\delta_C F_C)Y\rangle .
    \label{eq:delta_Skin_start_appC}
\end{equation}
Using Duhamel's formula and moving \(e^{s\mathcal{A}_C}\) to the first slot by
the formal adjoint gives
\begin{align}
    \delta_C S_\Delta^{\rm kin}
    &=
    -
    \frac12
    \int_0^1 ds\,
    \left\langle
        e^{s\mathcal{A}_C^\dagger}\Delta,
        B'(C)\delta C\,\Box_\Delta
        e^{(1-s)\mathcal{A}_C}Y
    \right\rangle .
    \label{eq:delta_Skin_adjoint_appC}
\end{align}
This is the adjoint identity underlying the local expression for
\(\Sigma_{\rm kin}\) in Sec.~\ref{sec:duhamel_variation}.  Explicitly,
\begin{equation}
    \delta_C S_\Delta^{\rm kin}
    =
    \int_{\mathcal{M}}d^4x\,\sqrt{-g}\,
    \Sigma_{\rm kin}\,\delta C,
    \label{eq:Sigma_kin_def_appC}
\end{equation}
with
\begin{equation}
    \Sigma_{\rm kin}
    =
    -
    \frac12
    B'(C)
    \int_0^1 ds\,
    \left[
        e^{s\mathcal{A}_C^\dagger}\Delta
    \right]_{\mu\nu}
    \left[
        \Box_\Delta
        e^{(1-s)\mathcal{A}_C}
        (\Box_\Delta\Delta)
    \right]^{\mu\nu}.
    \label{eq:Sigma_kin_appC}
\end{equation}

\subsection{Adjoint form of the memory-field variation}
\label{subsec:memory_variation_identity_appC}

At fixed \(C\) and fixed metric, the variation of the kinetic term with respect
to \(\Delta_{\mu\nu}\) is
\begin{equation}
    \delta_\Delta S_\Delta^{\rm kin}
    =
    -
    \frac12
    \langle \delta\Delta,F_C\Box_\Delta\Delta\rangle
    -
    \frac12
    \langle \Delta,F_C\Box_\Delta\delta\Delta\rangle .
    \label{eq:delta_Delta_Skin_appC}
\end{equation}
Using the formal adjoint and the self-adjointness of \(\Box_\Delta\),
\begin{equation}
    \langle \Delta,F_C\Box_\Delta\delta\Delta\rangle
    =
    \langle \Box_\Delta F_C^\dagger\Delta,\delta\Delta\rangle .
    \label{eq:Delta_variation_adjoint_appC}
\end{equation}
Therefore
\begin{equation}
    \delta_\Delta S_\Delta^{\rm kin}
    =
    -
    \frac12
    \left\langle
        \delta\Delta,
        F_C\Box_\Delta\Delta
        +
        \Box_\Delta F_C^\dagger\Delta
    \right\rangle .
    \label{eq:Delta_variation_final_appC}
\end{equation}
This identity gives the kinetic part of the memory-field equation in
Sec.~\ref{sec:memory_field_equation}.  In the constant-\(C\) limit,
\(F_C^\dagger=F_C\) and \([F_C,\Box_\Delta]=0\), so the expression reduces to
the familiar fixed-scale self-adjoint form.

\subsection{Domain and boundary assumptions}
\label{subsec:operator_domain_assumptions_appC}

The identities above are used under the regularity and boundary assumptions
stated in Sec.~\ref{subsec:regularity_assumptions_section2}.  In particular,
the fields are assumed smooth enough for the displayed operator compositions
to be meaningful, and boundary terms generated by integrations by parts are
assumed to vanish or to be cancelled by appropriate boundary data.  The
exponential \(F_C=\exp(\mathcal{A}_C)\) is treated as an analytic operator on
the chosen domain of symmetric tensor fields.

A full spectral analysis of \(\mathcal{A}_C\), including questions of global
well-posedness and pole structure around specific backgrounds, is separate from
the variational consistency result established in the main text.

\input{bib.tex}

\end{document}

%% file: bib.tex
\bibliographystyle{apsrev4-2}
\bibliography{references}